\documentclass[smallextended]{svjour3} 

\usepackage{caption}
\usepackage{booktabs}
\usepackage{latexsym}
\usepackage{graphicx}
\usepackage[titletoc,toc]{appendix}
\usepackage{gensymb}
\usepackage{amsmath}
\usepackage{cite}
\usepackage{url}
\begin{document}

\title{Focusing effect of bent GaAs crystals for $\gamma$-ray Laue
lenses: Monte Carlo and experimental results}

\author{E.~Virgilli, F.~Frontera, P.~Rosati, E.~Bonnini, E.~Buffagni, C.~Ferrari, 
 J.B.~Stephen, E.~Caroli, N.~Auricchio, A.~Basili, S.~Silvestri}

\institute{E.~Virgilli \and F.~Frontera \and P.~Rosati\at
              Department of Physics, University of Ferrara, Via Saragat 1/c, 44122 Ferrara, Italy
           \and
           E.~Bonnini \and E.~Buffagni \and C.~Ferrari\at
	   CNR-IMEM Institute, Parco Area delle Scienze 37/A, 43124 Parma, Italy
	   \and
	   N.~Auricchio \and A.~Basili \and E.~Caroli \and S.~Silvestri \and J.~Stephen \at
	   INAF/IASF  via Piero Gobetti, 101 - 40129 Bologna, Italy\\
           }

\date{Received: date / Accepted: date}

\authorrunning{Diffraction and focusing effect of GaAs bent crystals}
\titlerunning{Virgilli et al.}
\maketitle

\begin{abstract}

We report on results of  observation of the focusing effect from the planes (220) of Gallium Arsenide 
(GaAs) crystals.
We have compared the experimental results with the simulations of the focusing capability of
GaAs tiles through a developed Monte Carlo. The GaAs tiles were bent using a lapping process developed at the
{\sc cnr/imem} - Parma (Italy) in the framework of the {\sc laue} project, funded by ASI, dedicated 
to build a broad band Laue lens prototype for astrophysical applications in the hard 
X-/soft $\gamma$-ray energy range (80-600 keV).
We present and discuss the results obtained from their characterization, mainly in 
terms of focusing capability.
Bent crystals will significantly increase the signal to noise ratio of a telescope based 
on a Laue lens, consequently leading to an unprecedented enhancement of sensitivity with respect to 
the present non focusing instrumentation.

\keywords{Laue lens \and astrophysics \and bent crystals \and hard X-ray telescopes \and 
X-ray diffraction.}

\end{abstract}




\section{Introduction}
\label{sec:intro}

In hard X-/soft $\gamma$-ray astrophysics there is a strong need to investigate new 
technologies and detection methods capable of overcoming the limits exhibited
by current operational telescopes, both in terms of sensitivity and angular resolution. 
In the near future, Laue lenses made from bent crystals will have a key 
role in the hard X-/soft $\gamma$-ray  astronomy thanks to the 
technological progress made in the engineering and in the material 
science fields. 
One of the most attractive properties 
of  the Laue lenses is the possibility to increase the collecting surface without 
increasing the detection area and consequently the detector noise.

To date, mosaic crystals have been used to build a number of prototype capable to demonstrate 
the real possibility to build Laue lenses.  Mosaic crystals are composed of small blocks, called 
crystallites, with a size approximately
varying from 1 to 100 $\mu$m. The crystallites are assumed to be perfect crystals. The lattice
orientation of these crystallites are distributed around a main direction typically 
following a Gaussian distribution.  

In the sixties, about 4000 NaCl mosaic crystals have been used 
to focus the broad band 20-130 keV \cite{Lindquist68}. The mosaicity of about 0.5$^\circ$
resulted in a very wide energy response for each crystal but with low diffractive power.
During the 1980s, the Argonne National Laboratory of the University of Chicago 
studied both mosaic and bent crystals and many different designs for the crystal lenses
also developing a full size crystal diffraction lens made to cover the energy range from 200 keV to 1 MeV. 
Later, in a project in collaboration with the CESR laboratory at the University of Toulouse,
a lens for astrophysics was built ~\cite{Smither95} and 
for the first time an astrophysical source was observed
(the Crab nebula) during a stratospheric flight (the CLAIRE project~\cite{Halloin04}).
More recently, in the project named {\sc HAXTEL} (Hard X-ray TELescope) small prototypes of Laue lenses 
have been built and successfully tested to demonstrate the capability to set a number of mosaic Cu crystals 
with an accurate positioning system in a reasonable short time capable to focus energy of $\sim$100 keV~\cite{Frontera07, Virgilli11a}.
However, although flat crystals have the advantage of a relatively easy production in large quantity with 
good reproducibility in terms of dimensions and mosaic spread, they have limits that can be summarized as follows:

\begin{itemize}
 \item even without taking into account the absorption in the crystals, their maximum reflectivity 
 is limited to 50$\%$ which is the result of a radiation equilibrium between  direct 
 and diffracted beam from the crystalline planes\cite{Zachariasen}.

 \item a flat mosaic crystal tile does not display a focusing effect. Its diffracted image in the focal plane
 mainly depends on the crystal size and on the crystal mosaic spread which results in a defocusing 
 effect~\cite{Lund2005}.
 
\end{itemize}

Since the sensitivity of a Laue lens
strongly depends on the photon distribution on the focal plane, crystals capable of
focusing the photons on a small PSF are very promising.
Bent crystals can overtake one or both the mentioned limits, depending on their particular nature.
Two curvatures, that are conceptually different,  can be present in a bent crystal.
A curvature of the planes perpendicular to the X-ray beam can be 
induced (external curvature also named focusing curvature).
Thanks to the external curvature, bent crystals can focus the radiation at the focal plane 
detector into an area that is much smaller than the crystal cross section itself. Such a focusing 
effect can be observed for crystals with both perfect or mosaic structure.
A secondary curvature can be present if the internal diffraction planes are 
also bent (such a secondary curvature can be induced by the external curvature). 
For bent perfect crystals have been demonstrated that the secondary curvature increases  
their efficiency at values higher than  the 50\% of the incident radiation \cite{Malgrange02} which 
is the theoretical insurmountable limit for
flat crystals. Also for bent mosaic crystals, there are evidences that the curvature of the 
diffracting planes of the crystallites provides an efficiency that is larger than that 
of the equivalent flat mosaic crystals\cite{ferrari13b}. 

Bent crystals can be produced in a variety of methods~\cite{Smither05}. Curved diffractive planes are 
obtained by growing a two-component crystal where the relative concentration of the two components changes 
as the crystal is grown. An example of two component crystal has been grown at the  IKZ - Berlin
where Si-Ge crystals were grown with starting concentration of 3-7$\%$ percent of Germanium~\cite{keitel99}.
Perfect Si crystals were bent by applying a thermal gradient to a single structure 
crystal~\cite{Smither05}. Nevertheless, the power required to keep the curvature stable for 
each crystal of an entire Laue lens is too high and impracticable.
Instead, a mechanical bending can be more practical for astrophysical applications~\cite{Kawata01} 
given that a self-standing curvature does not require additional power.
Self-standing Si and Ge perfect crystals were produced and characterized at Sensor and 
Semiconductor Laboratory (Ferrara, Italy) by bending through mechanical grooving of one of
their surfaces~\cite{Bellucci11}. Although a good uniformity is achieved, 
the grooving process causes damage in the crystal. 
As a consequence, the crystal itself suffer from a high mechanical fragility and
the diffraction efficiency is limited to 50$\%$ if the diffraction 
occurs in the layers beside and beneath the grooves \cite{Camattari14}.
A self-standing curvature obtained by {\sc cnr/imem} - Parma~\cite{Buffagni11} relies on a controlled 
mechanical damaging on one surface of the sample. The procedure introduces defects in a superficial 
layer of a few microns, providing a highly compressive strain 
responsible for the convexity appearing on the worked side.

For the {\sc laue} project~\cite{virgilli13}, devoted to the building of Laue lenses for space astronomy in the 
hard X-ray energy band 
(80--600 keV), self standing bent crystals are being used. The tiles
made of Gallium Arsenide (220)~\cite{ferrari14, buffagni15} and Germanium (111)~\cite{camattari2012} have 
been bent using the two described method of lapping 
procedure and surface grooves, respectively. As the Laue lens has been
designed with a focal length of 20 m, the crystals have been prepared with a
curvature radius of 40 m.

In this paper we show the capability of a sample of GaAs mosaic crystals to strongly focus 
the radiation at the correct distance from the tile, if they are bent with the proper curvature radius. 
We also have studied the experimental set-up with a geometrical/analytic approach
and with Monte Carlo simulations. Both the cases of point like and extended X-ray source 
of radiation and their effect on the photon distribution on the focal plane detector 
are discussed and compared with the  results obtained with the laboratory beam line.

We show our capability to describe the diffraction process through the simulations  
in case of source of radiation placed  at a finite distance. The experimental 
results are in excellent agreement  with the expectations. Such a validation give 
us the confidence that the simulations made with a source of radiation 
placed at infinitive distance (astrophysical case) are also correct.

%

\section{A comparison between flat and bent crystals: Monte Carlo simulations}

All the simulations presented in this paper have been made  with the {\sc LLL} (Laue Lens Library) software 
developed by the
{\sc laue} project team. The code is developed for a complete 
description of Laue lenses made of mosaic or perfect crystals, with flat or bent structure, 
for a source  of radiation placed at either a finite distance
or infinitely far from the target.
The software is based on a ray tracer that takes into account the 
diffraction laws, the absorption caused by crystals of different materials and by the 
lens frame and allows the spatial distribution of
the photons at a given distance from the crystal after the X-ray diffraction to
be determined.
Different geometries of the lens have been implemented, as it is possible 
to select between 
single crystals, rings of crystals, sectors (petals) or entire lenses made 
of crystals set in concentric rings or in spiral configuration.
The software assumes an ideal focal plane X-ray imager detector (whose size
can be adapted to the particular application) with spectral capability.
The ray tracer and the detector are integrated in a unique software platform which has a single input 
front-end (where the crystal properties, the space between the crystals, the lens focal length, the 
energy pass-band must be defined) providing the physical configuration of the simulated lens (total 
weight of the lens, spatial dimensions, total geometric area) and its scientific features
(output energy pass-band, photon distribution on the focal plane, effective area, sensitivity).

\begin{figure}[!t]
   \begin{center}
   \includegraphics[scale=0.43]{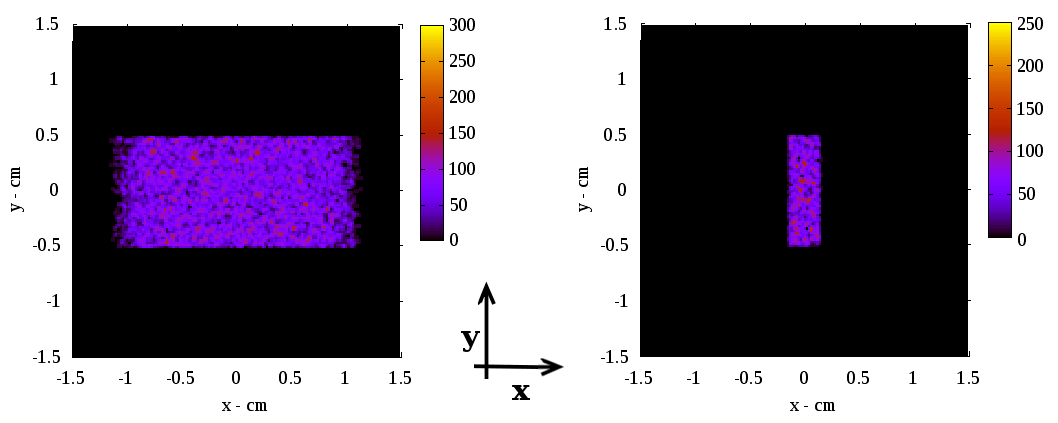}
     \caption{\footnotesize{Simulation of the diffracted image of a mosaic (15 arc seconds mosaicity) GaAs (220) crystal 
     tile with incident beam size of 20 $\times$ 10 mm$^2$ and source of radiation placed at infinite distance 
     from the target. {\it Left:} the case of a flat tile. {\it Right:} the case of a bent tile with 40 m curvature 
     radius. In both cases the detector is placed at 20 m from the target. The color scale represents the number of photons,
     for an impinging beam of 10$^6$ photons.}}
     \label{comparison:inf:flat:bent}
 \end{center}
 \end{figure}
 
When a infinitely far source impinges over a flat perfect crystal, the radiation encounters the diffractive planes 
with the same Bragg angle, as the source
of radiation has no divergence. The result is that the diffracted image has the
same size as the crystal cross section itself. In the case of a mosaic crystals, the mosaic spread 
results in a defocusing effect
and the size of the diffracted image is given by the convolution between the crystal size itself 
and its mosaicity. Figure~\ref{comparison:inf:flat:bent}(left) shows the diffracted signal from the (220) planes 
of a flat mosaic GaAs crystal with mosaicity of 15 arc seconds, cross section 30 $\times$ 10 mm$^2$, 2 mm thick, 
when an X-ray beam of dimension 20 $\times$ 10 mm$^2$ impinges on it. The beam size was set smaller than the crystal cross 
section itself to avoid undesired border effects.  
The simulated x-rays have a uniform distribution of 10$^6$ photons in the energy range 149.7 - 150.3 keV which is roughly 
the pass-band offered by the defined mosaicity when the Bragg angle is set to diffract 
the centroid energy of 150 keV.

\begin{figure}[!t]
   \begin{center}
   \includegraphics[scale=0.15]{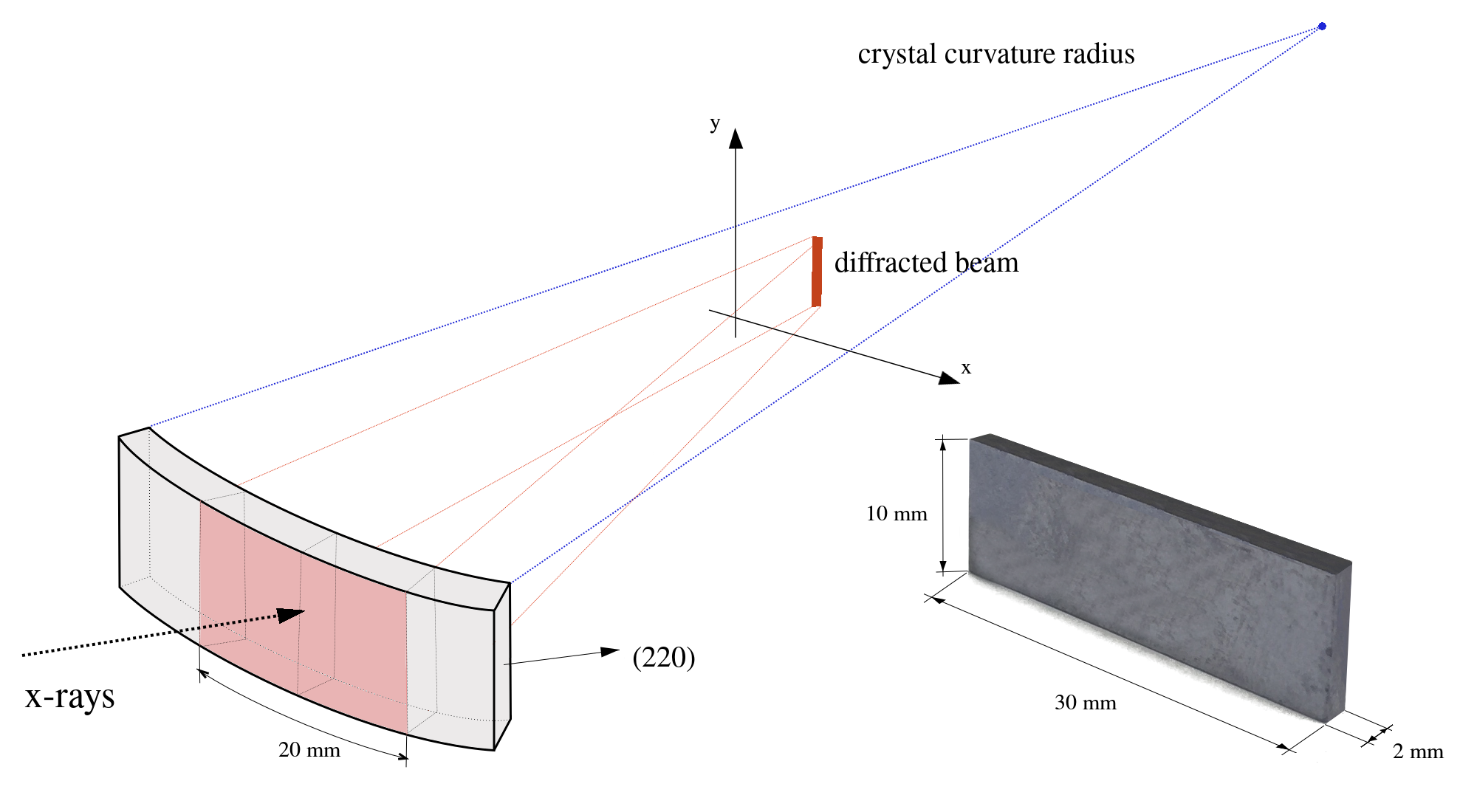}
     \caption{\footnotesize{{\it Left:} Sketch of a bent GaAs crystal and the principles of the diffraction from the (220) planes.  
     The crystal dimensions and the diffraction planes are also indicated.  The curvature radius of 40 meters 
     is also indicated and overstated for sake of clarity. The dimensions of the X-ray beam are also drawn over the 
     crystal cross section. 
      {\it Right:} Picture of a GaAs crystal tile used for the experimental run.}}
     \label{bent_real_and drawing}
 \end{center}
 \end{figure}

A comparison is made with the simulation of the diffracted beam from the (220) planes of a bent GaAs mosaic crystal 
having the same dimensions and mosaic spread, 
but with 40 m curvature radius (Fig.~\ref{comparison:inf:flat:bent}, right). As exemplified in the sketch 
of Fig.~\ref{bent_real_and drawing}, the spot dimension is strongly reduced by the focusing effect dictated by 
the curvature of the crystal in the curvature direction, even if the mosaic defocusing is still present. 
The result is that for a bent crystal the space distribution of the 
photons in the focal plane  along the {\it x} direction is
smaller than the crystal length. Differently, no focusing effect is expected 
in the other direction, thus along the {\it y} profile the width of the diffracted image has 
the same size of the crystal.
It is also worth noting that for a bent crystal the energy 
pass-band is much wider than that shown by a flat GaAs and and for the particular case, it turns out to be
in the range  148.5 keV - 151.5 keV.

%
%

\section{Effects of the source divergence}
\label{div_effect}

The focusing effect shown in Fig.~\ref{comparison:inf:flat:bent} (right panel) is obtained for the 
case of a source of radiation placed at infinite distance. Unfortunately, in the laboratory 
the source of radiation is at a finite distance from the crystal and the focusing effect
as obtained from an astrophysical source cannot be directly observed. It can be demonstrated 
that the combination of the sample curvature 
and the divergence of the incoming beam
makes the radiation focus at a different focal distance, depending on the separation 
between the radiation source and the crystal.

\begin{figure}[!h]
\begin{center}
\includegraphics[scale=0.18, angle=0]{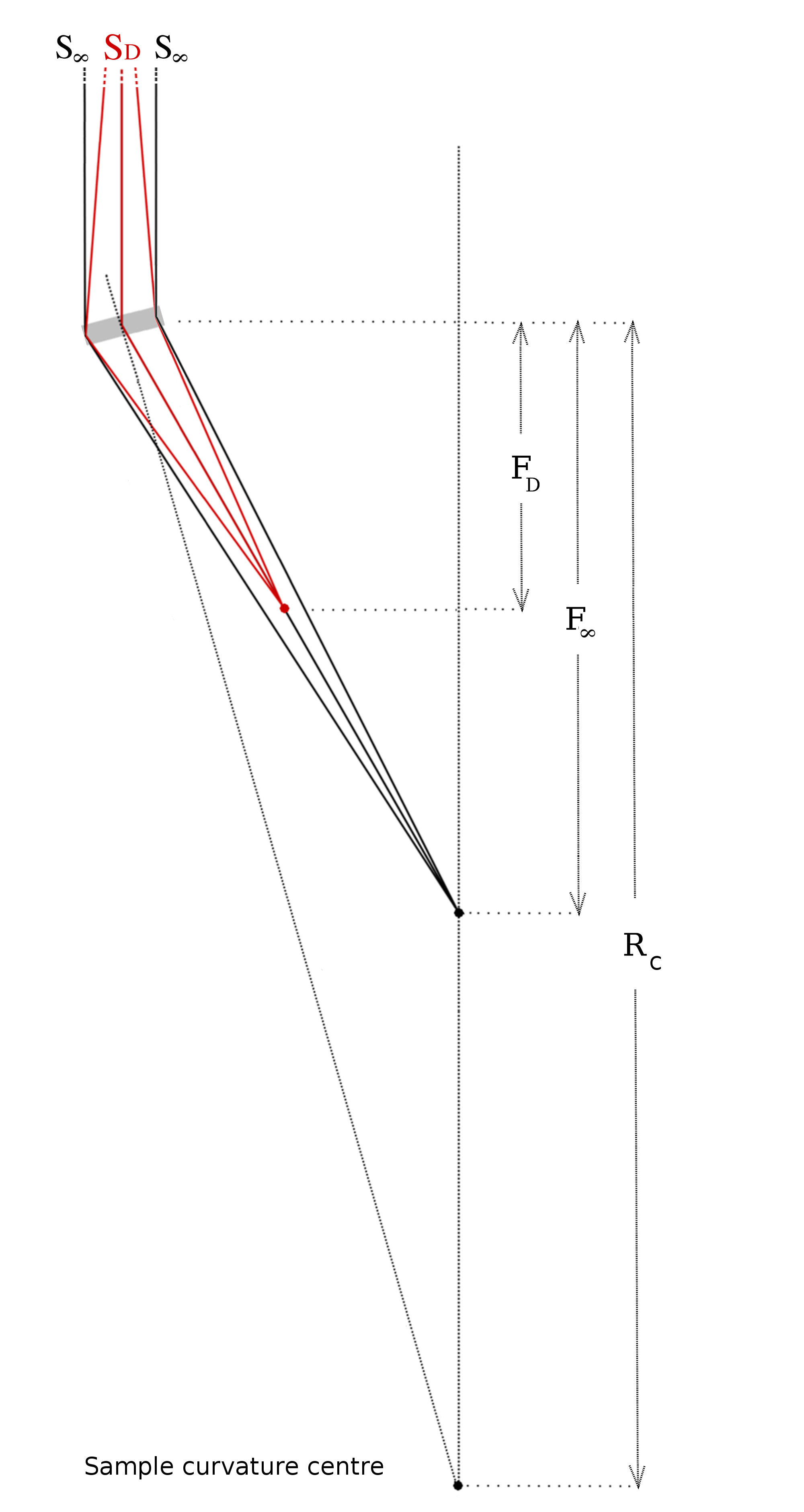}
\caption{\footnotesize{Effect of the divergence of the beam in the focusing effect for a 
bent crystal with a given curvature radius.}}
\label{ferrariexplainationa}
\end{center}
\end{figure}

\begin{figure}[!h]
\begin{center}
\includegraphics[scale=0.18, angle=0]{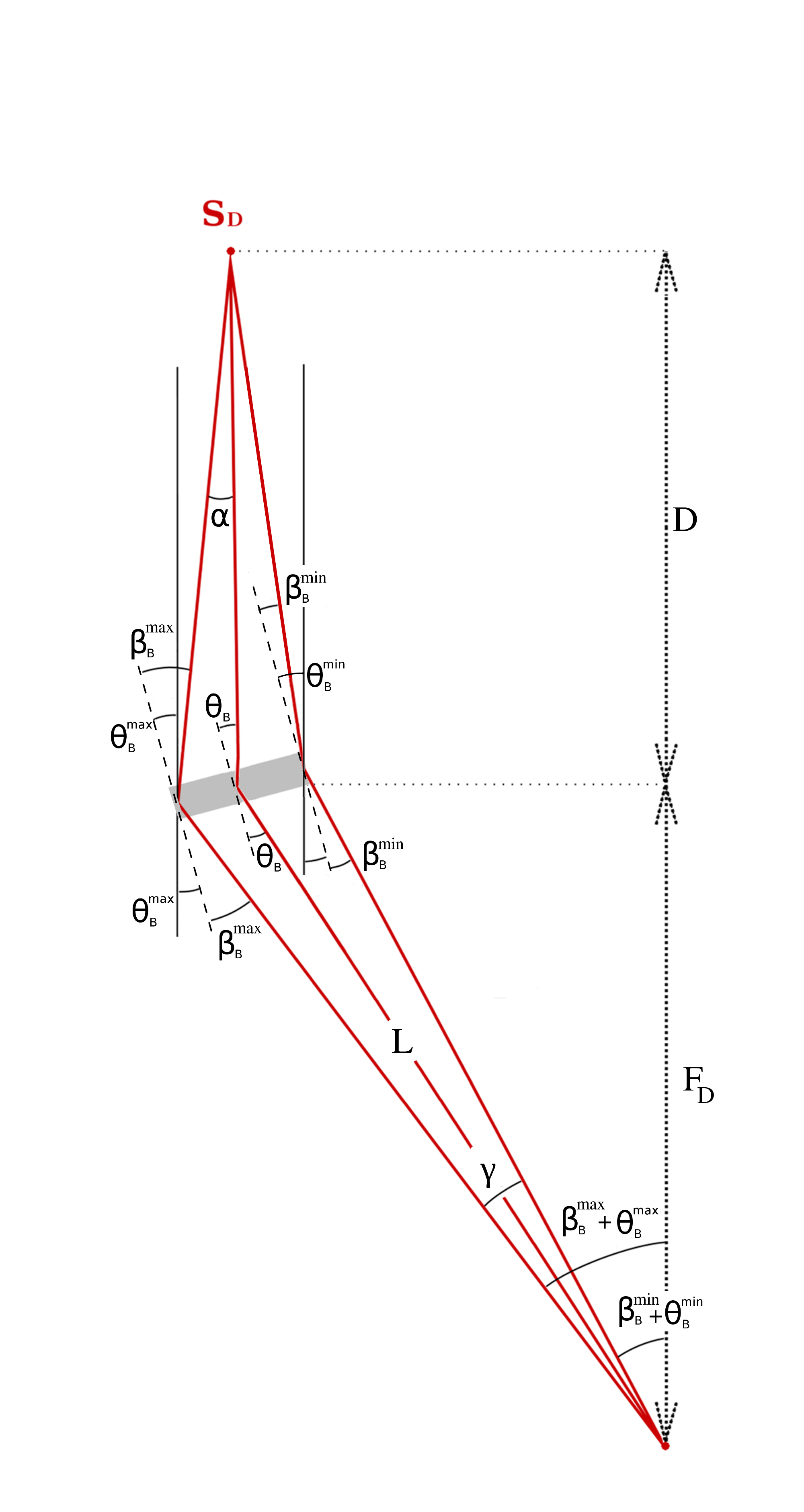}
\caption{\footnotesize{Detail of Fig.~\ref{ferrariexplainationa} where only the diffracted beam from a finite 
source is described.}}
\label{ferrariexplainationb}
\end{center}
\end{figure}

A geometrical explanation of the effect can be deduced. Let us assume a crystal sample 
with cross section $l$ $\times$ $s$ (bent along the direction $l$) irradiated by a source placed at infinite 
distance $S_\infty$ with no divergence. If 
$\theta_B$ is the corresponding Bragg angle for the radiation 
impinging at the center of the sample (Figs.~\ref{ferrariexplainationa} and \ref{ferrariexplainationb}), 
the Bragg angles at the extreme 
points of the sample along its length $l$ depend on $l$ and on the curvature radius of the sample $R_c$:

\begin{equation}
\theta_B^{min} = \theta_B - \frac{l}{2~R_c}
\end{equation} 
\begin{equation*}
\theta_B^{max} = \theta_B + \frac{l}{2~R_c}
\end{equation*} 

Under these conditions, the diffracted radiation is focused at the point $F_\infty$ = $R_c$/2 (see Fig.~\ref{ferrariexplainationa}). 
On the other hand, if the source of radiation $S_D$
is placed at a finite distance $D$ (see Fig.~\ref{ferrariexplainationb}), the Bragg angle $\beta$ at the center of 
the crystal does not change with respect to the case of an infinitely far source ($\beta_B$ = $\theta_B$)  while, due 
to the combined effect of 
the divergence angle $2\alpha$ and the curvature angle $\Delta$, at the extreme points 
of the sample the diffraction angles are defined as follows:

\begin{equation}
\beta_B^{min} = \theta_B^{min} - \alpha = \theta_B - \frac{l}{2~R_c} - \frac{l}{2~D}
\end{equation} 
\begin{equation*}
\beta_B^{max} = \theta_B^{max} + \alpha = \theta_B + \frac{l}{2~R_c} + \frac{l}{2~D}
\end{equation*}

and the radiation is focused at a different focal distance 
$F_D$. To calculate the value of $F_D$ it is convenient
to consider the following relations (see Fig.~\ref{ferrariexplainationb}):

\begin{equation*}
L ~ \gamma = l  ~~~~~~~~~~~~~~~~~  L~cos~2\theta_B = F_D 
\end{equation*}

That together give: 

\begin{equation*}
F_D = \frac{l ~ cos~2\theta_B}{\gamma} \nonumber 
\end{equation*}

in which the value of $\gamma$ is defined as:

\begin{equation}
\gamma = ( \beta_B^{max} + \theta_B^{max} ) - ( \beta_B^{min} + \theta_B^{min} ) = \frac{l}{D}+ \frac{2~l}{R_c} = \frac{l}{D}+ \frac{l}{F_\infty} 
\label{eq:gamma}
\end{equation} 

The value of the new focal distance $F_D$ is then determined as a function of the 
sample curvature and the source distance and does not depend on the
beam size over the target. 

%
%

For small diffraction angles ($cos{2\theta_B} \approx 1$) the relation becomes: 

\begin{equation}
\frac{1}{F_D}  \approx \frac{1}{D} +  \frac{2}{R_c}
\label{eq:fd2}
\end{equation} 

Instead, with no approximations, the focal distance $F_D$ depends on energy and its
variation is $\Delta F_D$ $\sim$ 12 mm $@$ 90 keV and $\sim$ 3 mm $@$ 200 keV.


\newpage
    
\newpage
 
This approach allows the value of the best focal distance to be determined, for a crystal
whose curvature radius is known.
Indeed the curvature of a sample can be independently estimated by using the method of the K$\alpha$ line. 
The method is based on the estimation of the tilt angle required to diffract the K$\alpha$ energy (59.20 keV)
along the crystal length. Being curved the crystal surface, the required angle changes along the crystal length 
and the curvature can be calculated from the slope of the linear relation between the crystal coordinate 
hit by the radiation and  the required  tilt angle providing the diffracted energy.
The method is extensively described in  
\cite{Liccardo12} where the achieved accuracy in the curvature radius estimation was within $\sim$2\%. 
In the case of the {\sc larix} facility \cite{loffredo03, Virgilli11a, larix} 
the source is at the distance distance $D$ = 26.40 $\pm$ 0.02~m from the sample, then for a crystal sample 
with a curvature radius of 40~m, the best focal distance is $F_D$ = 11.39~m.

\begin{figure}[!h]
\begin{center}
\includegraphics[scale=0.3, angle=0]{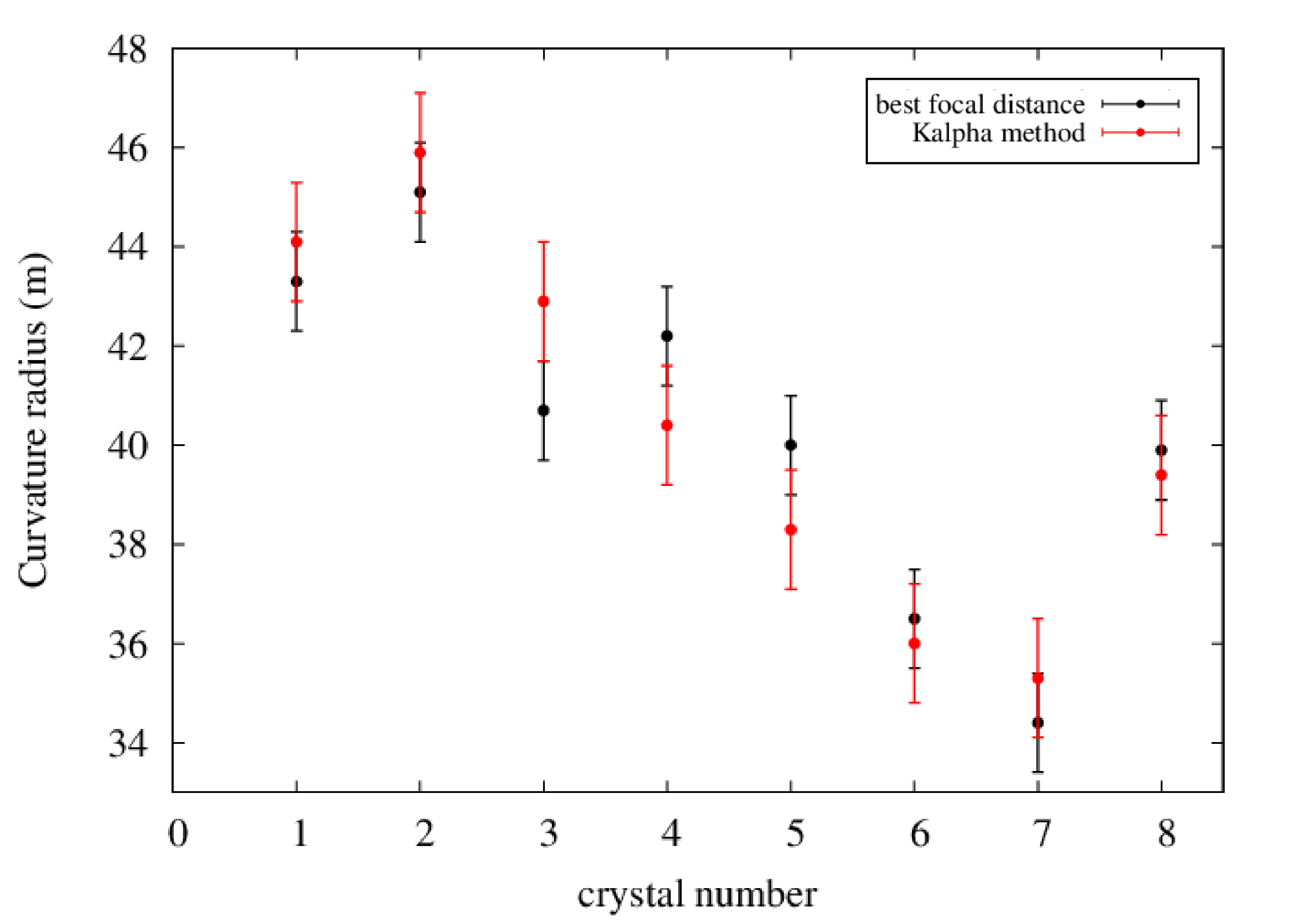}
\caption{\footnotesize{Estimation of the curvature radius for a sample of 8 GaAs crystals bent using the surface lapping 
method {\sc cnr/imem} - Parma. Red points correspond to the curvature estimation through the K$\alpha$ method, 
black points correspond to the estimation via the best focal distance which minimizes the {\sc psf fwhm}.}}
\label{radiuscaculation}
\end{center}
\end{figure}

Alternatively, for a fixed value of $D$, it is possible to calculate the curvature radius 
of a sample by simply measuring the value of $F_D$ that is the distance at which the width of 
the {\sc psf} is minimized. The method of the distance can be validated if compared with the 
described K$\alpha$ method and the experimental comparison is shown in 
in Fig.~\ref{radiuscaculation} where the values of the curvature radius have been estimated for 
a sample of 8 crystals. The methods are in good agreement within the uncertainties.

\begin{figure}[!h]
   \begin{center}
     \includegraphics[scale=0.24]{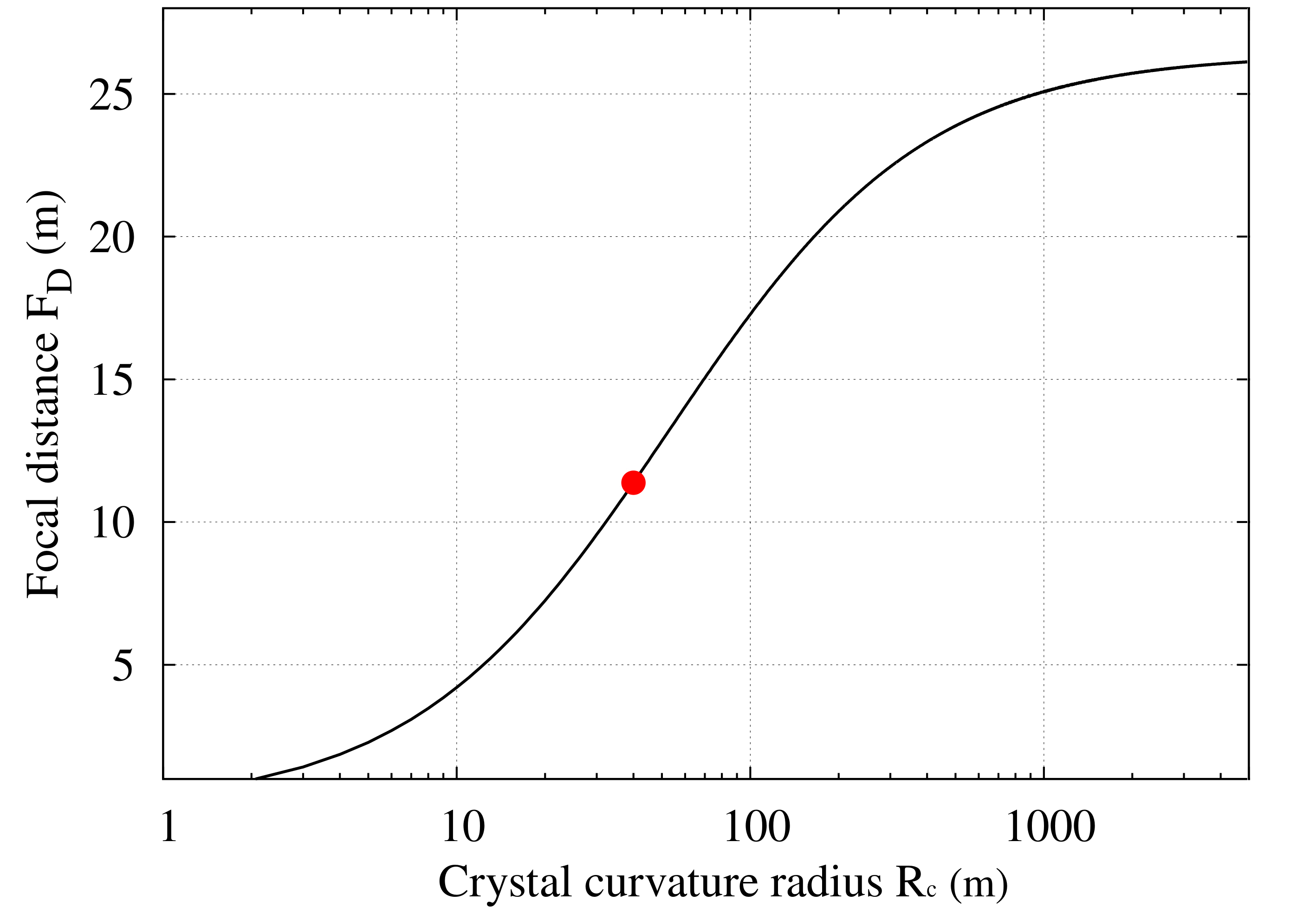}
     \includegraphics[scale=0.24]{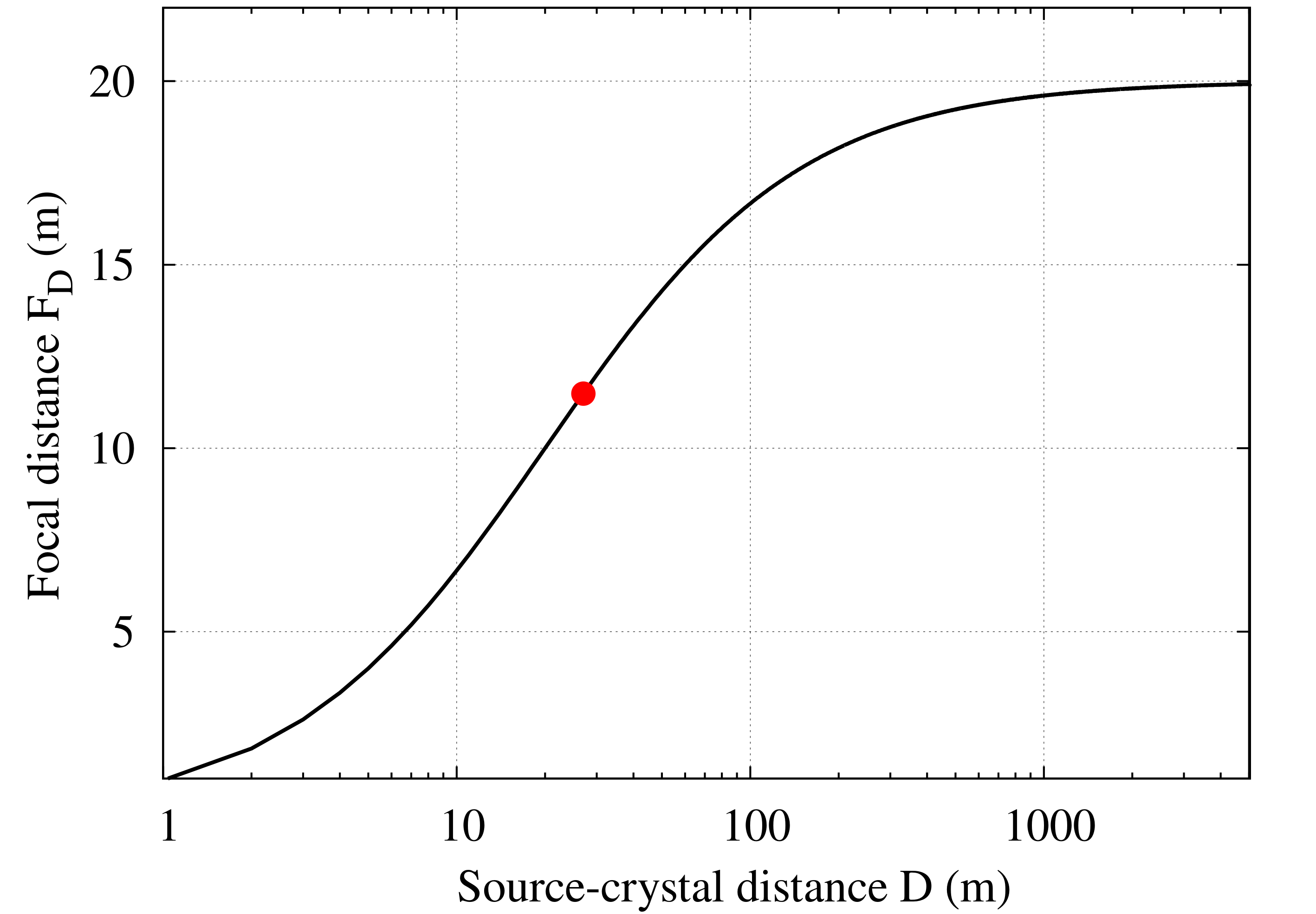}
      \end{center}
     \caption{\footnotesize{{\it Left}: the focal distance $F_D$ is shown as 
     function of the tile curvature radius for a fixed distance $D$ between source and crystal ($D$ = 26.40~m).
     At the limit of flat crystal (R $\rightarrow$~$\infty$) the focusing 
     distance is $F_D$ = 26.40~m equal to the source-target distance $D$ = 26.40~m.
     The red point ($F_D$ = 11.40~m) corresponds to the case 
     of  40~m curvature radius. 
     {\it Right}: the focal distance $F_D$ is shown as function of the distance between source and crystal, 
     for a fixed value of the crystal curvature radius ($R_c$ = 40 m). If the source is placed at infinite distance 
     ($D$ $\rightarrow$~$\infty$) it turns out that the focal distance $F_D$  is 20~m, as expected. The red dot represents the 
     $F_D$ = 11.40~m which is obtained with the {\sc larix} setup of source-crystal distance $D$ = 26.40~m. }} 
     \label{differentRorDifferentD}
 \end{figure}

It is interesting to discuss the expectations of Eq.~\ref{eq:fd2} in the limiting cases of a flat 
crystal or a source placed at infinite distance. In the case of a flat crystal ($R_c$~$\rightarrow$~$\infty$) for a 
source at finite distance (e.g., $D$ = 26.40~m), it turns out that the focusing position occurs 
at $F_D$ = $D$, as expected (see Fig.~\ref{differentRorDifferentD}, left panel). 
On the other hand, if the crystal curvature radius 
is fixed, and $D~\rightarrow~\infty$, then $F_D$ = $F$, also as expected.
The dependence of $F_D$ on $D$ for a fixed curvature radius is shown in Fig.~\ref{differentRorDifferentD}, right panel.

\begin{figure}[!h]
   \begin{center}
    \includegraphics[scale=0.36]{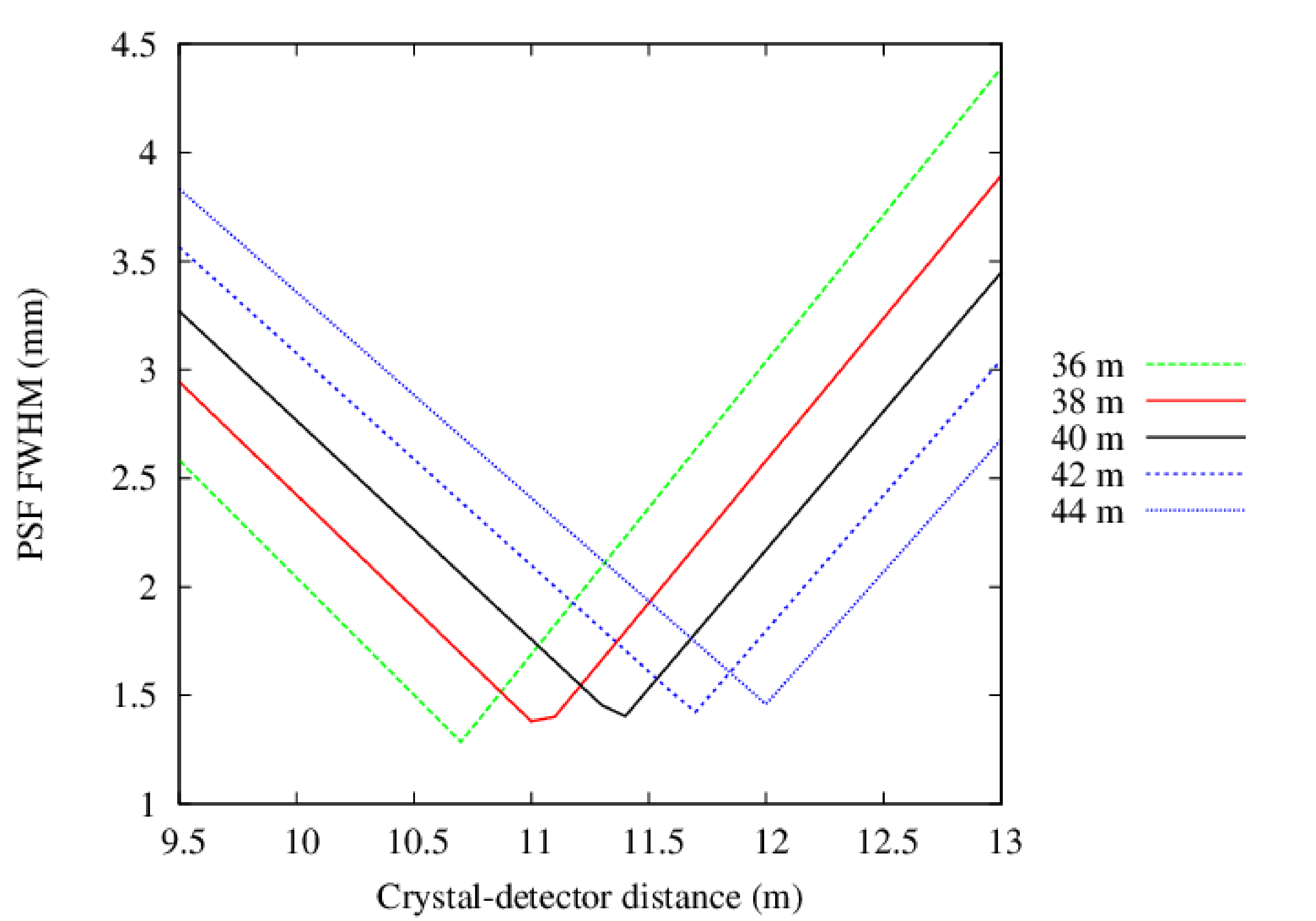}
     \caption{\footnotesize{{\sc psf} {\sc fwhm} plotted as function of the detector-crystal distance 
     for different values of the crystal curvature radius.}}
     \label{40meters}
 \end{center}
 \end{figure}

In addition to the ray tracer, a geometrical method have been
developed to derive the Full Width at Half Maximum ({\sc fwhm})
of the Point Spread Function ({\sc psf}) as a function of the crystal-detector distance 
for a given value of the curvature radius of the crystal. In this approach the determination 
of the {\sc fwhm} is obtained by considering the mutual distance 
between the diffracted  X-rays coming from the two extreme points of the crystal at different
distances, taking  into account the crystal intrinsic angular spread, the beam divergence, and the 
curvature effect along the $l$ direction. 
The results based on this analytic/geometric calculation, as a function of the crystal-detector 
distance and for various crystal curvature radii, are shown in Fig.~\ref{40meters}.

\section{Effects of the extended dimensions of the X-ray source}
\label{sec:mc}

The ray tracer have been also used to estimate the effect of an extended 
source of radiation on the photon distribution on the focal plane and to compare 
the result with that achievable with a point-like source of radiation.
Indeed, a real X-ray tube source has a finite dimension that cannot be
neglected. In the {\sc larix} facility, the X-ray tube employed for the measurements  is equipped 
with two interchangeable targets where the radiation is generated (foci):

\begin{itemize}
 \item small source ({\sc ss}): size 0.35 $\times$ 0.45~mm$^2$ (800 W); 
 \item large source ({\sc ls}): size 0.85 $\times$ 0.95~mm$^2$ (1800 W).
\end{itemize}

The Monte Carlo simulations allow us to derive the dependence of the {\sc psf} {\it x} profile 
(i.e., along the diffraction direction) on the source size. A 
uniform photon distribution in the 100-300 keV energy range was 
assumed, with the source kept at a distance of 26.40 m from the crystal, while the 
source size was varied with different size from 0 to 3~mm.

\begin{figure}[!h]
   \begin{center}
    \includegraphics[scale=0.32]{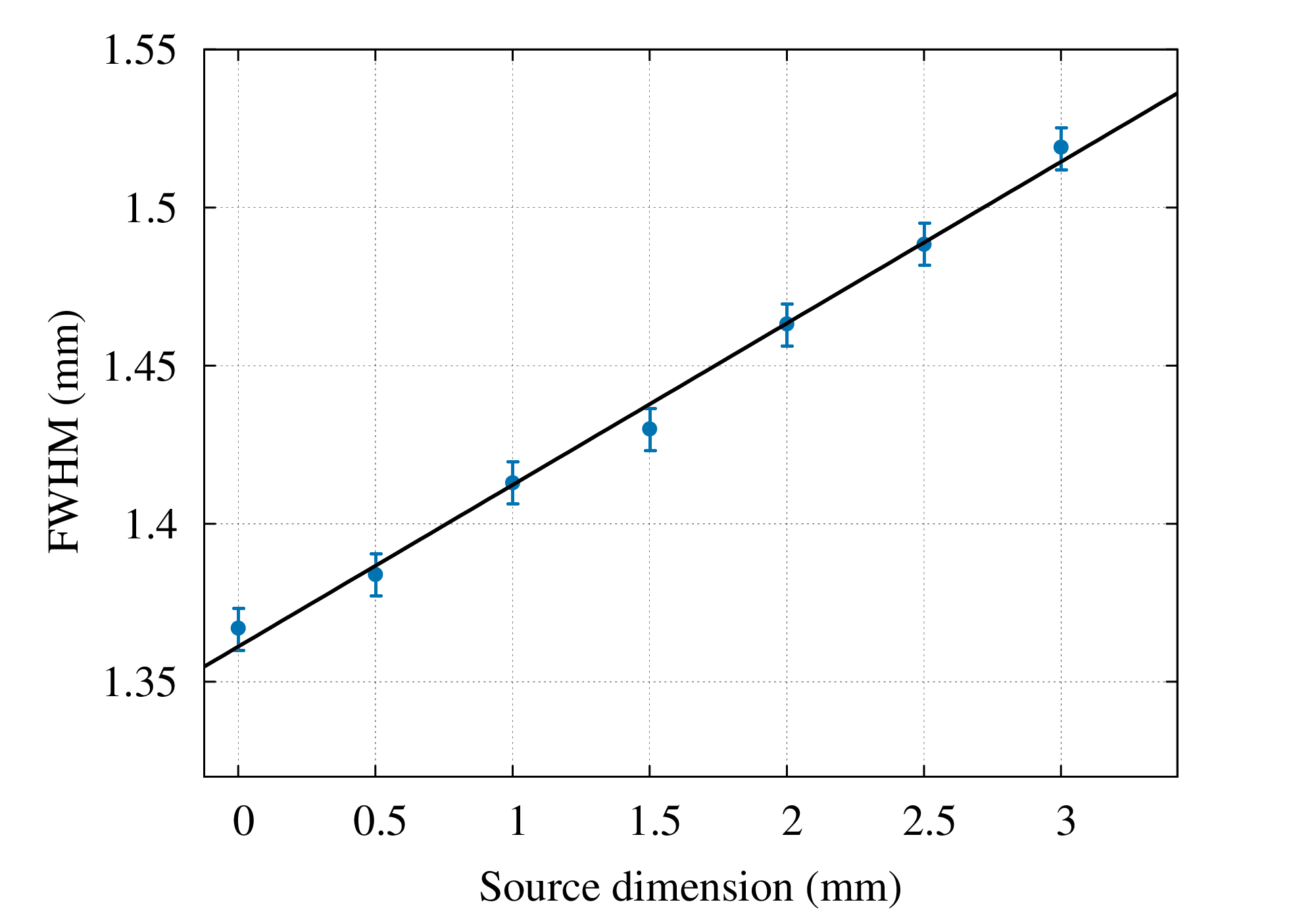}
     \caption{\footnotesize{Simulation of the  {\sc fwhm} of the diffracted beam for different values 
     of the X-ray beam size (blue points), from point-like source (source dimension = 0)
     to a source with diameter of 3~mm. The black line represents the linear best fit.}}
     \label{focalspot}
 \end{center}
 \end{figure}

It is evident from Fig.~\ref{focalspot} that there is a linear dependence
between the dimension of the extended source in the calculation of the 
{\sc psf} {\sc fwhm}. In the ideal case of a point like source the 
{\sc psf} {\sc fwhm} results to be 1.365 mm which only depends on the 
mosaicity of the sample, assumed for our purposes to be $\sigma$ = 20 arc seconds.
Instead, for the given value of the mosaicity, the minimum {\sc psf} {\sc fwhm} using the equipped 
{\sc ss} and {\sc ls} results {\sc fwhm$_{SS}$} $\sim$ 1.38
and {\sc fwhm$_{LS}$} $\sim$ 1.41, respectively.

\section{Effect of the finite distance of the source}
\label{mc}

Monte Carlo simulations also provide information on the {\sc psf}
produced by a crystal when the radiation source is placed at a finite distance from the crystal.
It is worth noting that a Laue lens for space applications
does not suffer of divergence effects, while, being the lens developed on ground, the source is 
at a finite  distance and the beam divergence cannot be neglected.

\begin{figure}[!h]
   \begin{center}
   \includegraphics[scale=0.42]{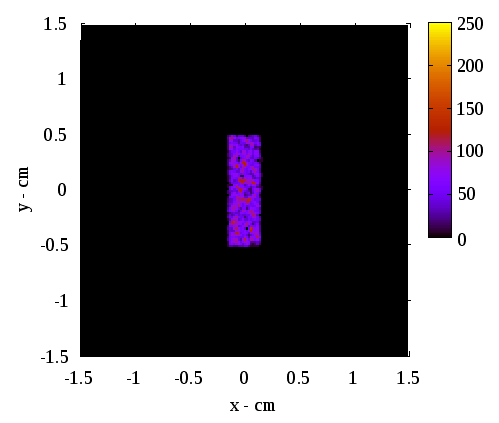}
   \includegraphics[scale=0.42]{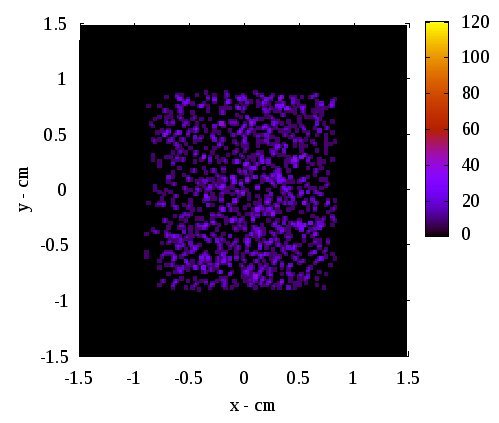}
   \includegraphics[scale=0.22]{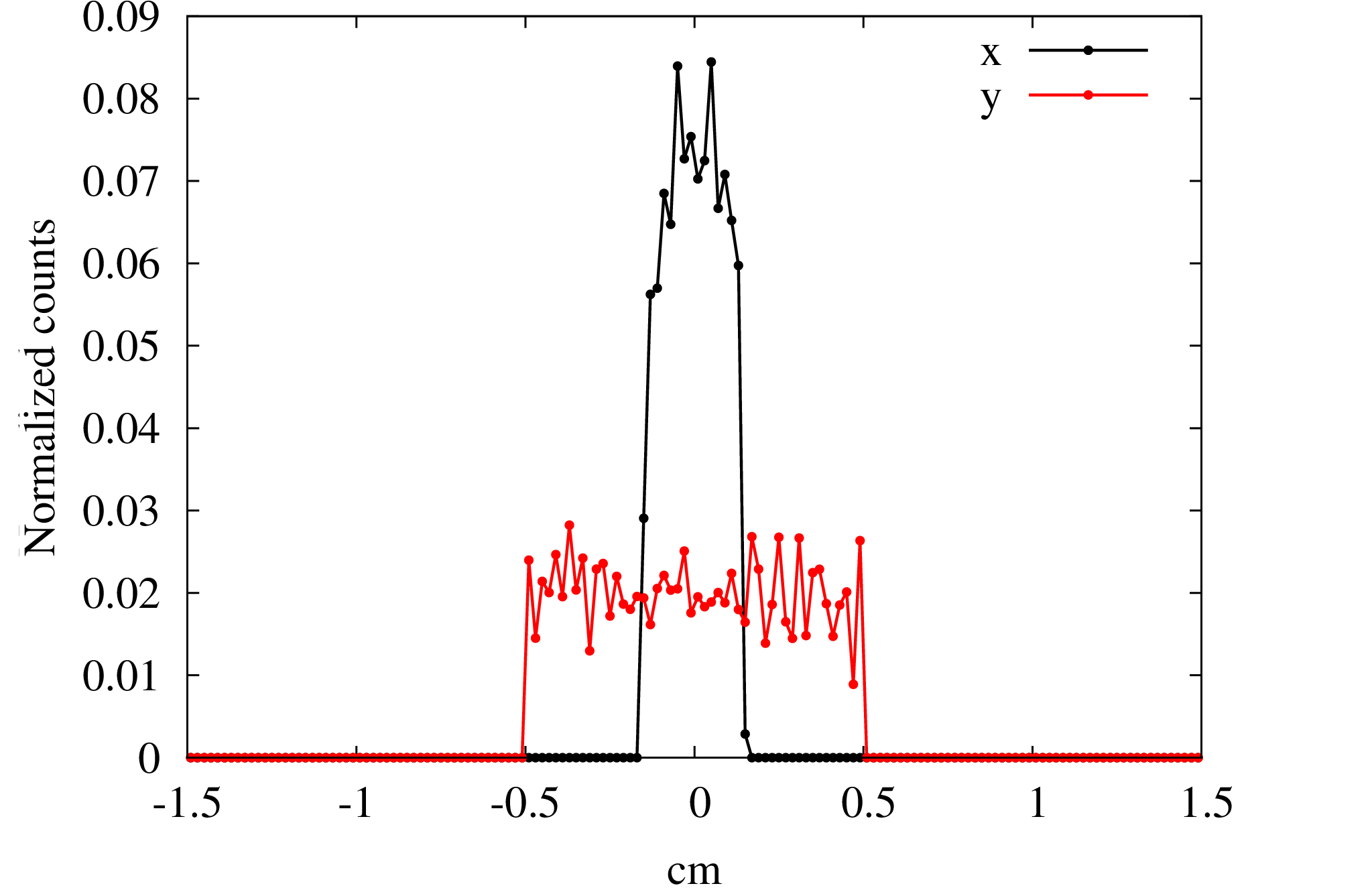}
   \includegraphics[scale=0.22]{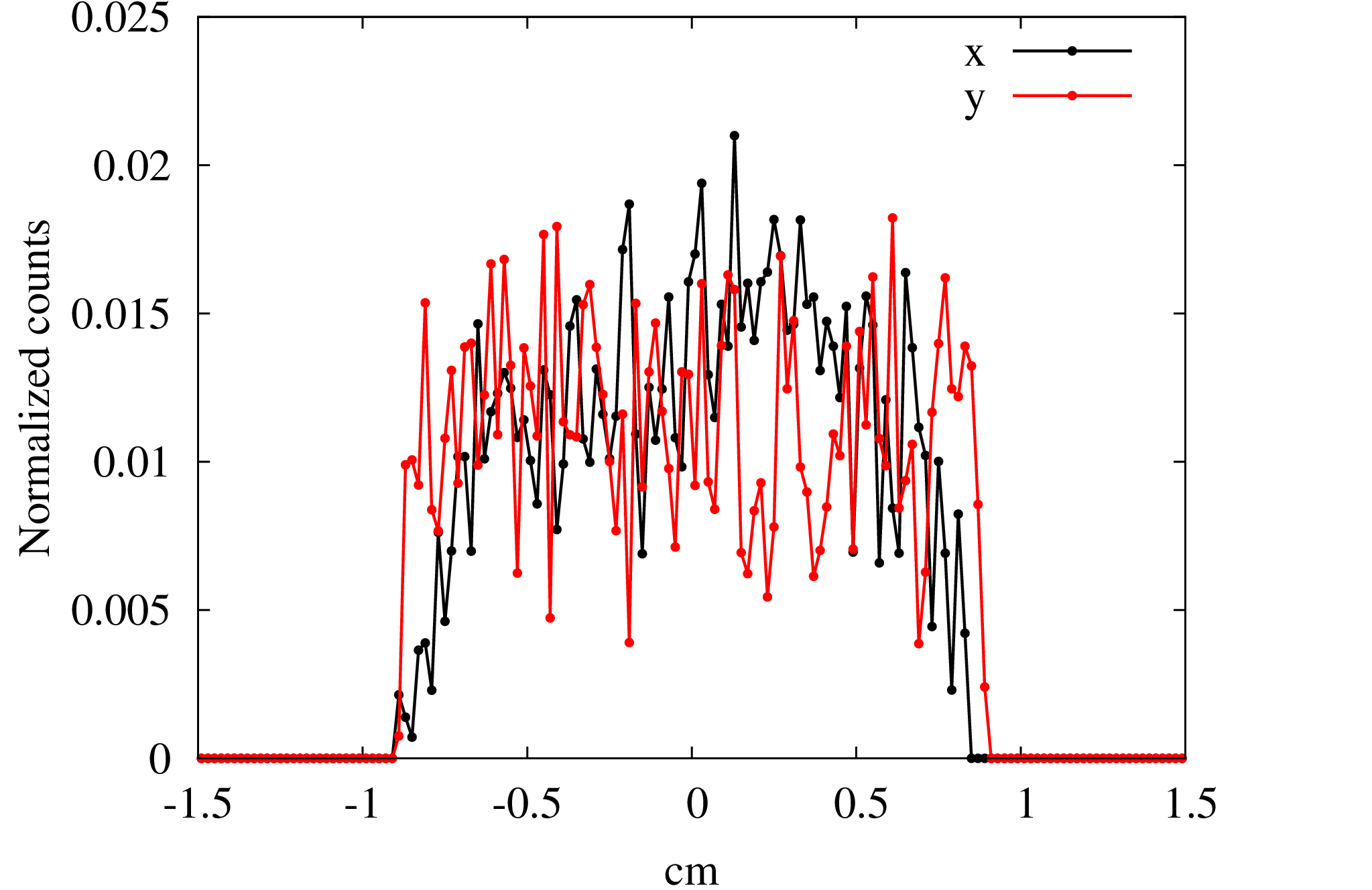}
     \caption{\footnotesize{Comparison between the diffracted images expected for a parallel beam 
    (left) and for a beam from a point like source placed at 26.40 meters from 
     the target (right) with the position sensitive detector placed at 20 m. In both cases a bent GaAs mosaic 
     crystal tile with a beam size impinging on it of 
     size 20 $\times$10 mm$^2$ have been considered.}}
     \label{diff:confronto_infinito_26m}
 \end{center}
 \end{figure}

Simulations have been made assuming a source at infinite distance and at a distance $D=26.40$~m. In 
the latter case, it is assumed that the radiation is emitted from the {\sc ss} of the X-ray tube. 
In both the cases the radiation impinges over an area of 20 $\times$ 10 mm$^2$ centered on the 
30 $\times$ 10 mm$^2$ GaAs (220) crystal with a curvature radius $R_c = 40$~m. 
The diffracted beam is observed at the nominal focal distance of 20~m.
The results are shown in Fig.~\ref{diff:confronto_infinito_26m} (top panels), together with the 
{\it x} and {\it y} distribution profiles (bottom panels).

%

The apexes and indices used in the following definition of {\sc fwhm} are self-explanatory. 
Along the {\it y}  direction, as expected, the focusing effect is not present and 
the divergence effect is only present in the case of a source placed 
at finite distance D. Under these conditions we get {\sc fwhm}$^y_D$ $\sim$ 17.5 mm
and {\sc fwhm}$^y_\infty$ $\sim$ 10.0 mm which is the crystal dimension itself along the {\it y} direction.  
Regarding the x direction, the value {\sc fwhm}$^x_\infty$ $\sim$ 2.9 mm is mainly related to the 
mosaic spread of the sample. Instead,  in the case of the source placed at 26.40 m, 
it drastically increases ({\sc fwhm}$^x_D$ $\sim$ 16.5 mm) due to a defocusing 
effect dictated by the combination of the beam divergence with the curvature of the sample.

 \begin{figure}[!t]
   \begin{center}
   \includegraphics[scale=0.49]{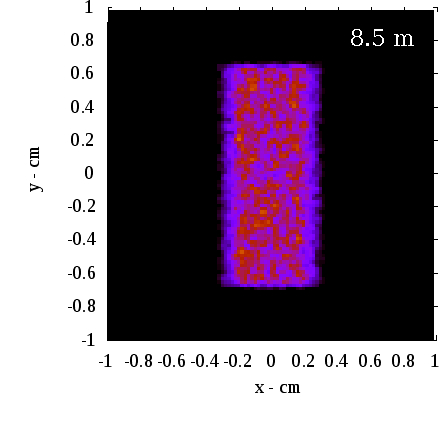}
   \includegraphics[scale=0.49]{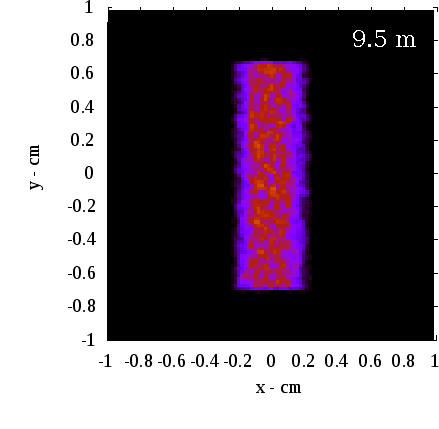}
   \includegraphics[scale=0.49]{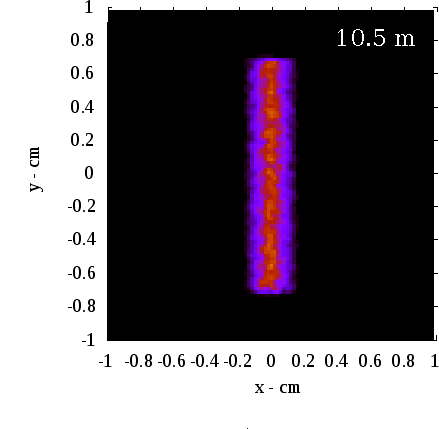}
   \includegraphics[scale=0.49]{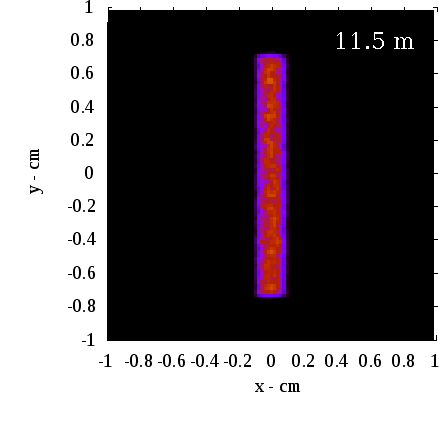}
   \includegraphics[scale=0.49]{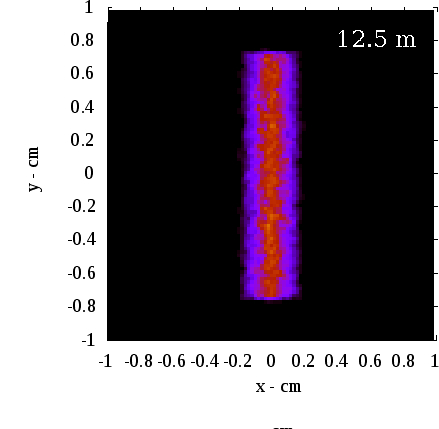}
   \includegraphics[scale=0.49]{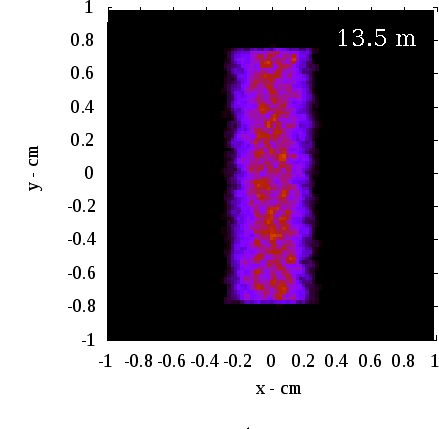}
     \caption{\footnotesize{Simulation of the diffracted image generated by a GaAs bent crystal
     with a beam size of 20 $\times$ 10 mm$^2$, observed at 8.5, 9.5, 10.5, 11.5, 
12.5, 13.5 m from the crystal (respectively from the left top to the right bottom).}}
     \label{image:diffracted}
 \end{center}
 \end{figure}

The defocusing effect observed in Fig.~\ref{diff:confronto_infinito_26m} (right panels) is expected
and can be removed if the diffracted beam is observed at a closer distance from the crystal.
Figure~\ref{image:diffracted} shows the spatial distribution of the signal when the diffracted photons are 
collected  at different distances from the sample (from 8.5 m to 13.5 m). The minimum 
{\sc fwhm} of the {\sc psf} is  obtained by collecting the photons at $\sim$ 11.4 m from the crystal, in agreement 
with the analytical  calculations (Eq.~\ref{eq:fd2}). 

\begin{figure}[!h]
   \begin{center}
   \includegraphics[scale=0.45]{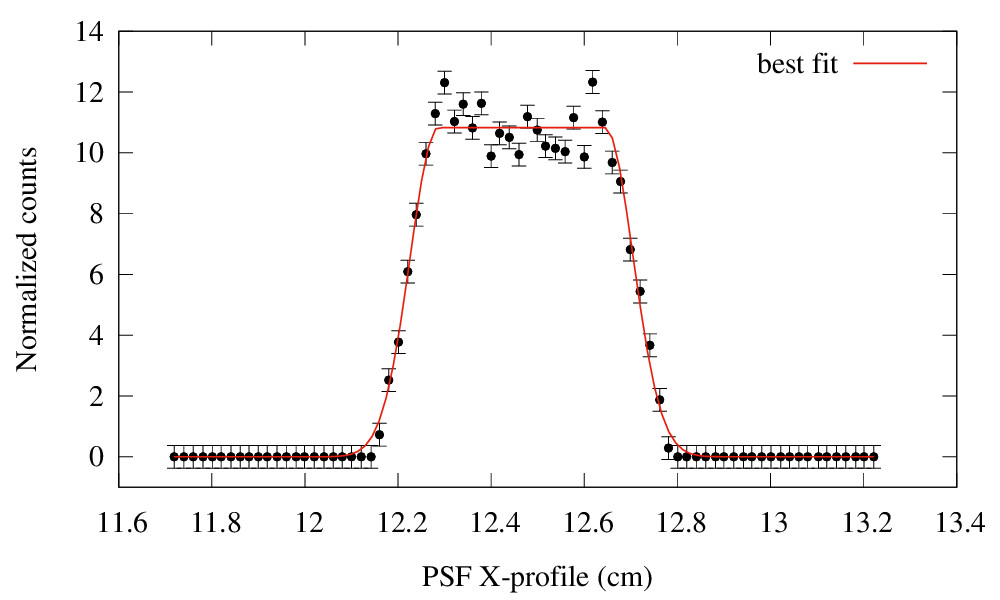}
   \includegraphics[scale=0.45]{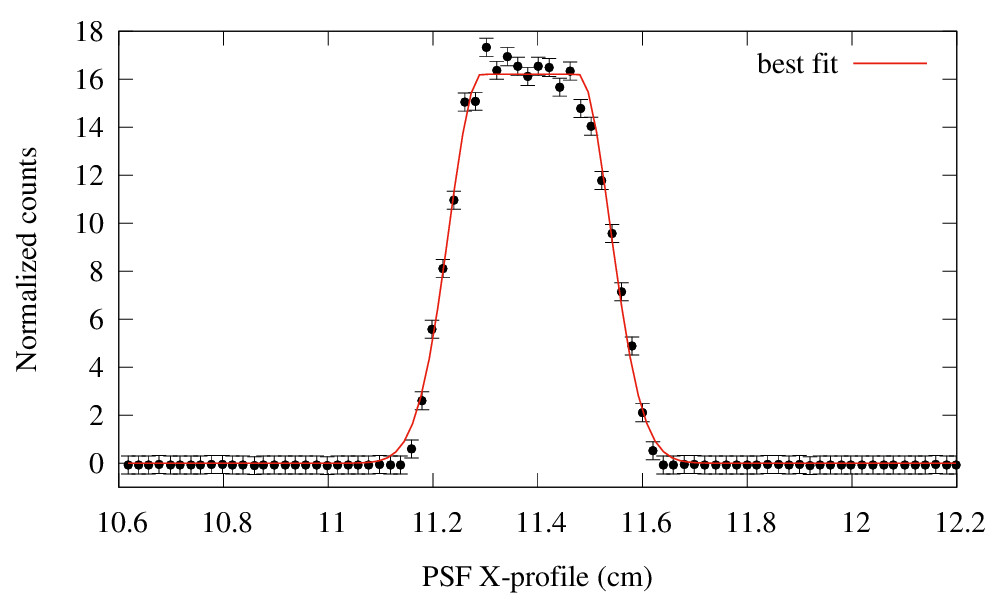}
   \includegraphics[scale=0.45]{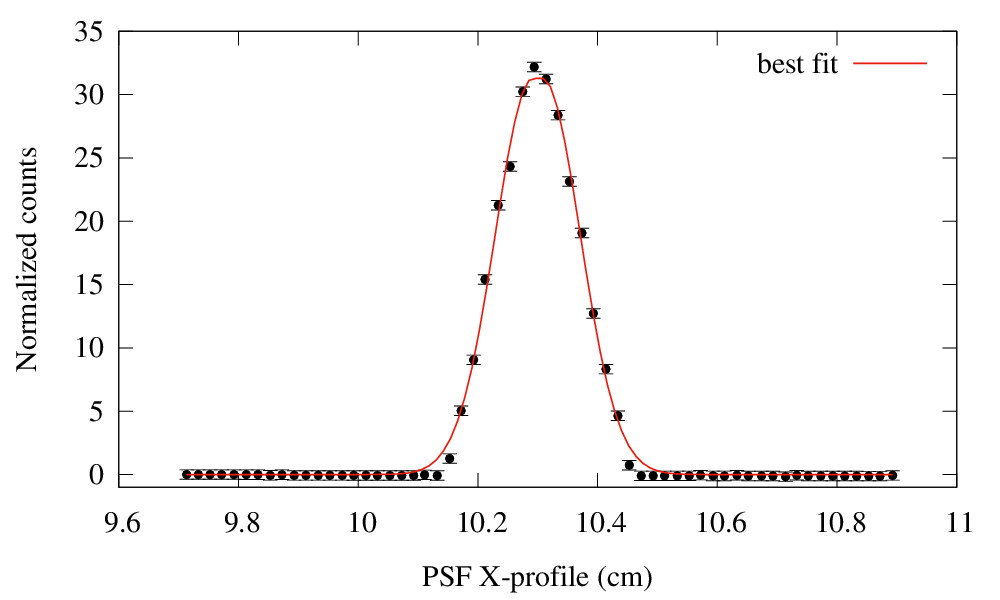}
   \includegraphics[scale=0.45]{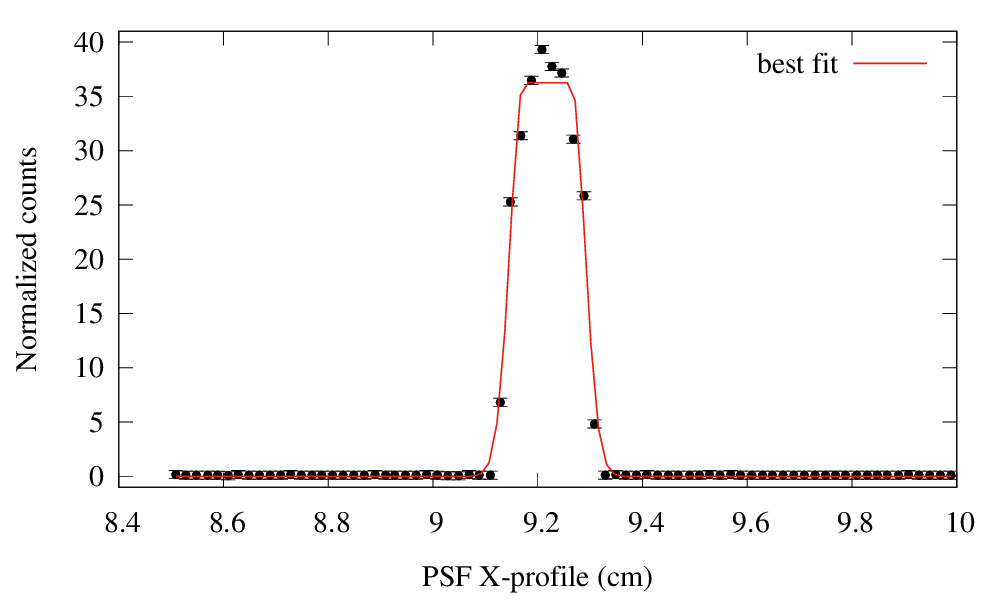}
   \includegraphics[scale=0.45]{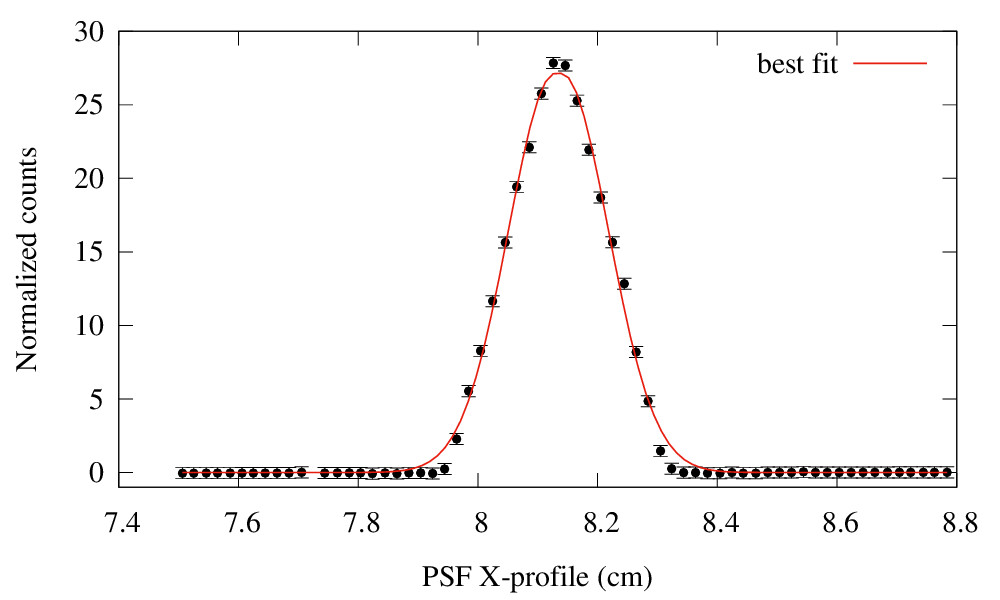}
   \includegraphics[scale=0.45]{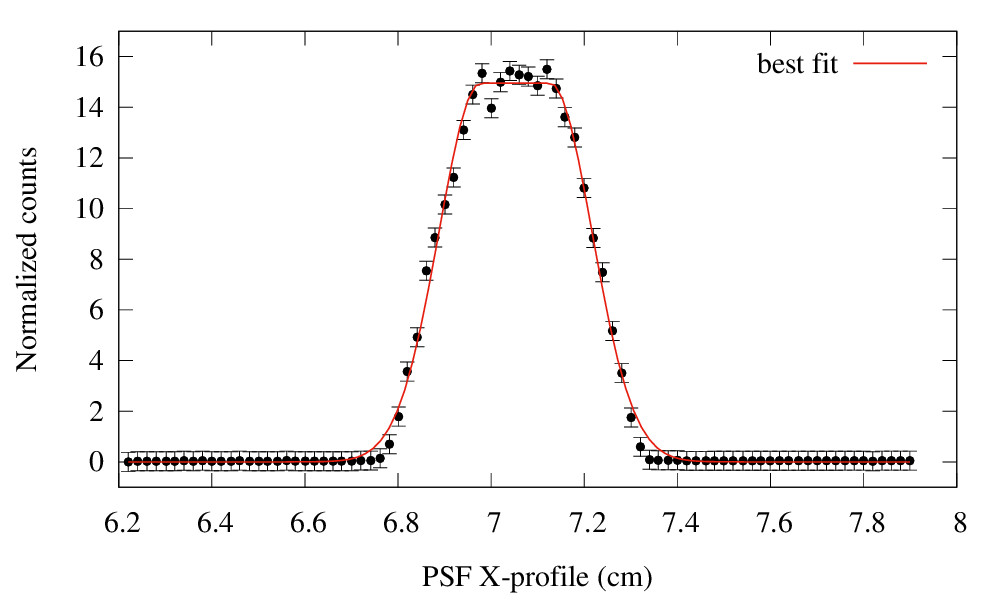}
     \caption{\footnotesize{ X-profile of the diffracted image generated by a GaAs bent crystal
     with a beam size of 20 $\times$ 10 mm$^2$ cross section observed at 8.5, 9.5, 10.5, 11.5, 12.5, 13.5 m 
from the crystal (respectively from the left top to the right bottom).}}
     \label{image:fit}
 \end{center}
 \end{figure}

In Fig.~\ref{image:fit} are also shown the {\it x} profiles of the images acquired at different distances
with the red line representing is the best fit function. A satisfactory function is represented by a convolution between
a Gaussian profile with a fixed value of {\sc fwhm} and a rectangular function with variable width.  
Within this explanation, the width of the Gaussian function represents the angular spread which is constant 
along the entire crystal length. In the experimental case the mosaicity of the sample was of $\sim$ 15 arc seconds.
When the diffracted image is acquired at the expected focal distance the spread of the photon distribution  
is comparable with the Gaussian spread, given that all the rays are 
converging almost in a point. Instead, out of the focus the Gaussian profile is uniformly spread into a 
larger area which is analytically described by the rectangular function, whose width depends on the 
distance of the detector from the best focal distance.

\newpage
 
\section{Experimental set-up}
\label{sec:setup}

\begin{figure}[!h]
   \begin{center}
   \includegraphics[scale=0.29]{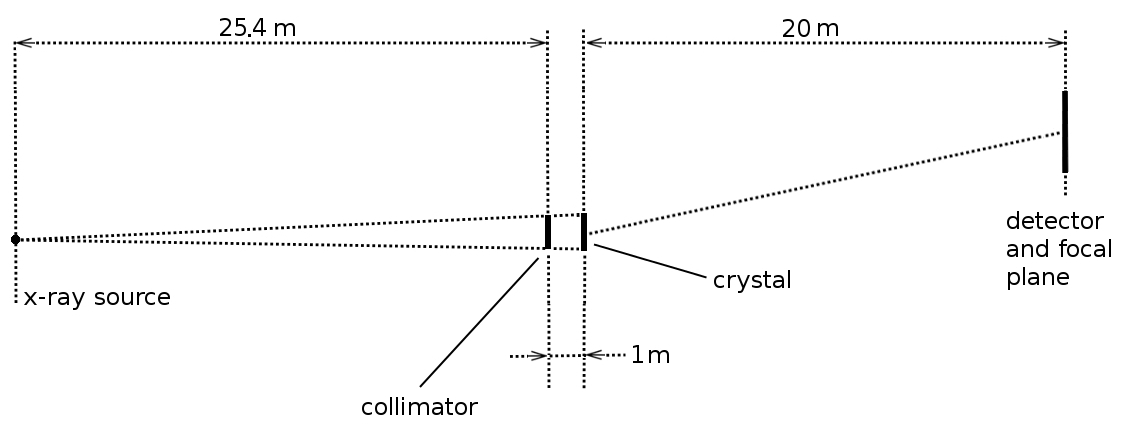}	
     \caption{\footnotesize{Sketch showing the set up (top view) and the distances between the subsystems.}} 
     \label{setup}
 \end{center}
 \end{figure}

The results of the simulations have been validated through an experimental run 
performed in the  {\sc larix} facility.
The {\sc larix} equipment has been extensively described elsewhere~\cite{Virgilli11b,virgilli13}. 
As portrayed in Fig.~\ref{setup}, the distance between 
the source and the target is 26.40 $\pm$ 0.02 m, while the distance between the collimator and crystal is 1 $\pm$ 0.02 m.
All the degrees of freedom of the subsystems are entirely motorized as 
required for the alignment procedures. The collimator blades can be also remotely adjusted in order to
set the vertical and the horizontal beam dimensions.

 \begin{figure}[!h]
   \begin{center}
    \includegraphics[scale=0.037]{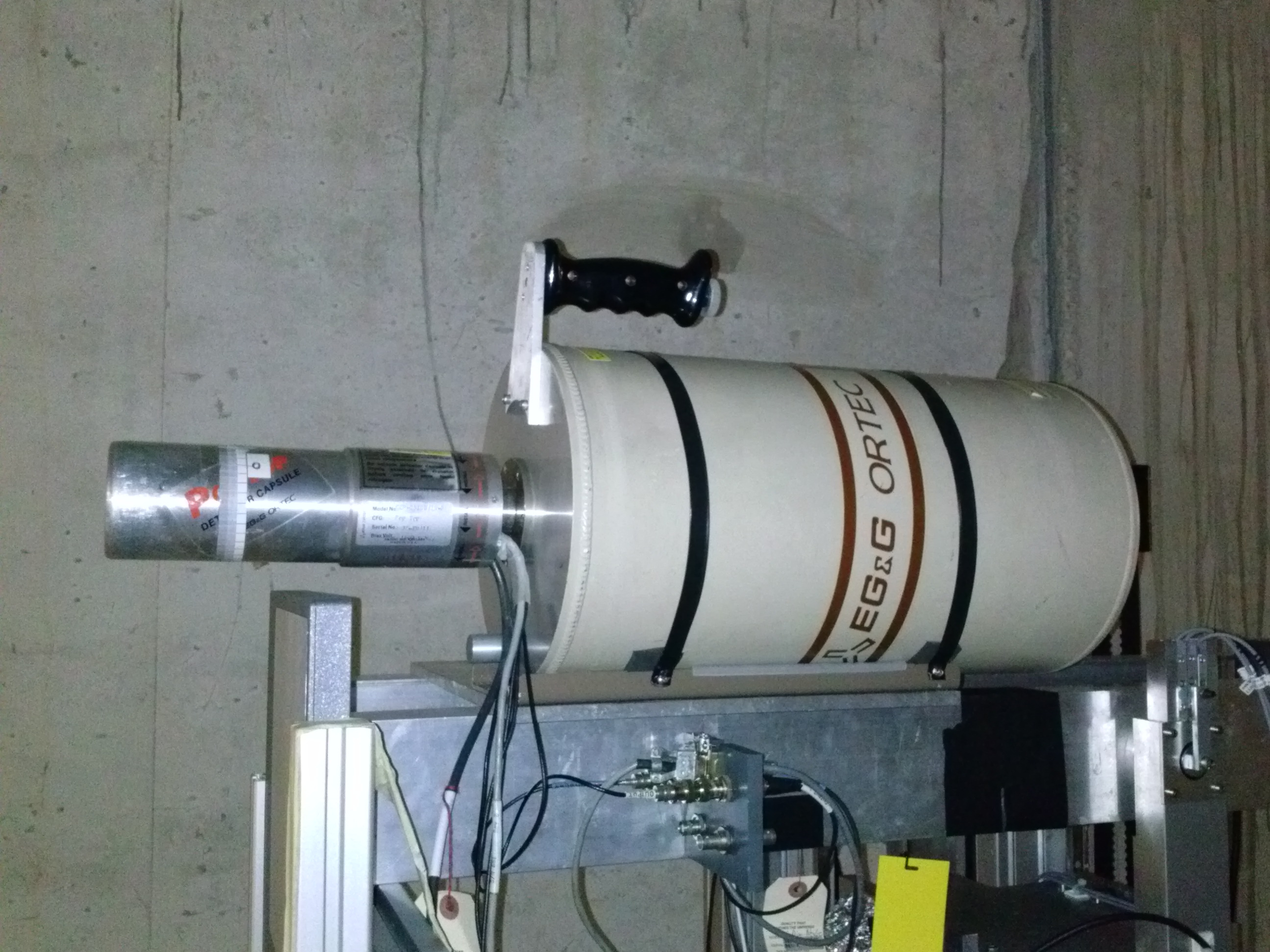}
    \includegraphics[scale=0.037]{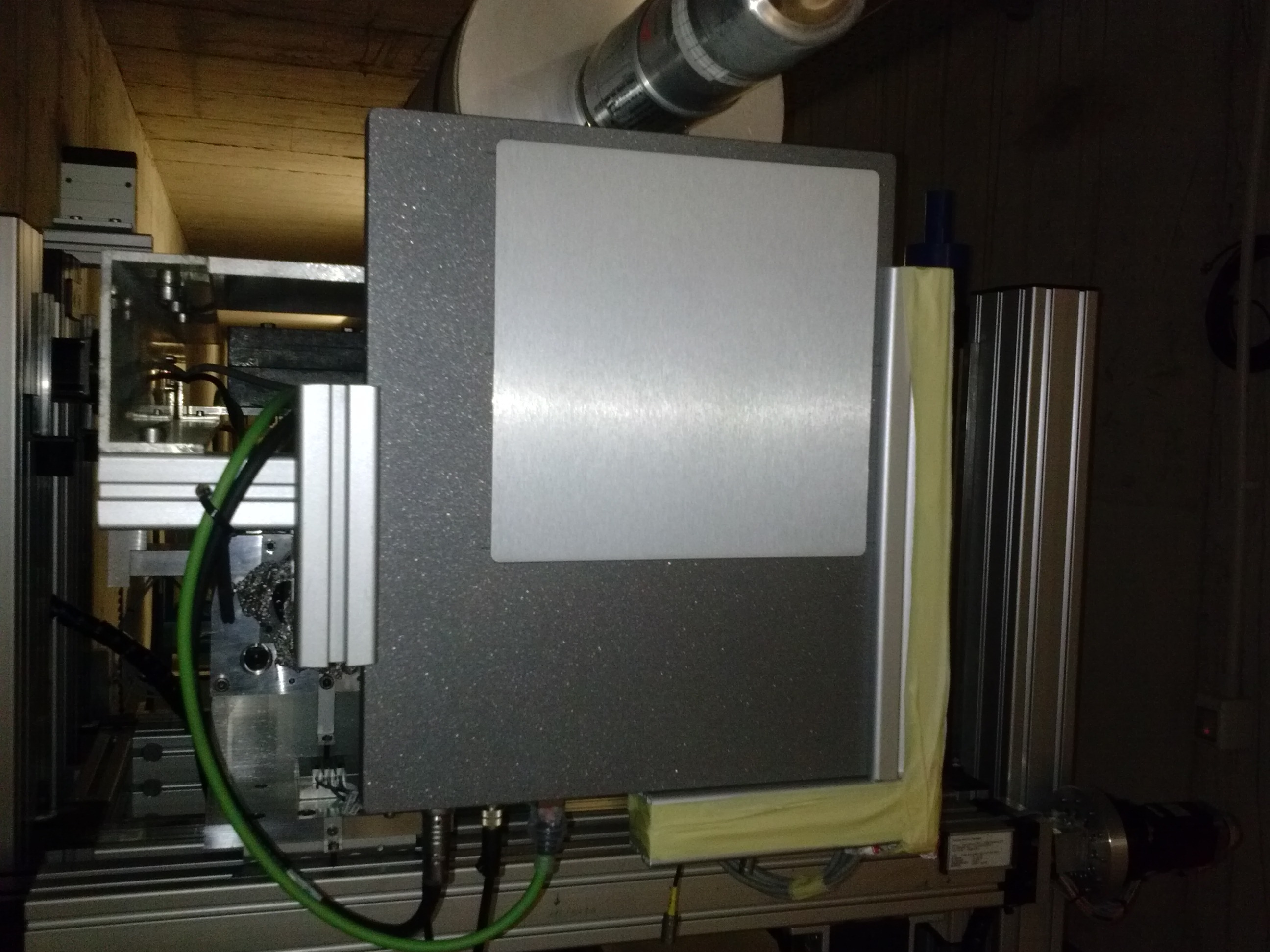}
     \caption{\footnotesize{The detectors used for the measurements in the  {\sc larix} facility. {\it Left:}
     HPGe spectrometer with the cooling system device also installed in the tunnel.
     {\it Right:} flat panel X-ray imaging system, installed on the detector carriage.}}
     \label{detectors}
\end{center}
\end{figure}

Two detectors have been used in the test campaign (Fig.~\ref{detectors}). For spectral 
analysis a portable High Purity Germanium Spectrometer (HPGe, resolution 500 eV $@$ 80 keV; 
550 eV $@$ 122 keV) with a  liquid nitrogen cooling system is available (left panel). A digital 
X-ray detector with a total collecting area of 20 $\times$ 20 cm$^2$ was used to acquire the 
diffracted images (right panel). The imager consists of a 800 $\mu$m thick CsI panel, directly deposited on a Si sensor operating as a 
two-dimensional photodiode array of 1024 $\times$ 1024 elements, providing a spatial 
resolution of 200 $\mu$m. Each frame has a maximum integration time of 17 seconds but multiple frames can 
be summed off-line in order to increase the total exposure time and the image quality.


\section{Test results}
\label{sec:results}

\begin{figure}[!h]
   \begin{center}
    \includegraphics[scale=0.3]{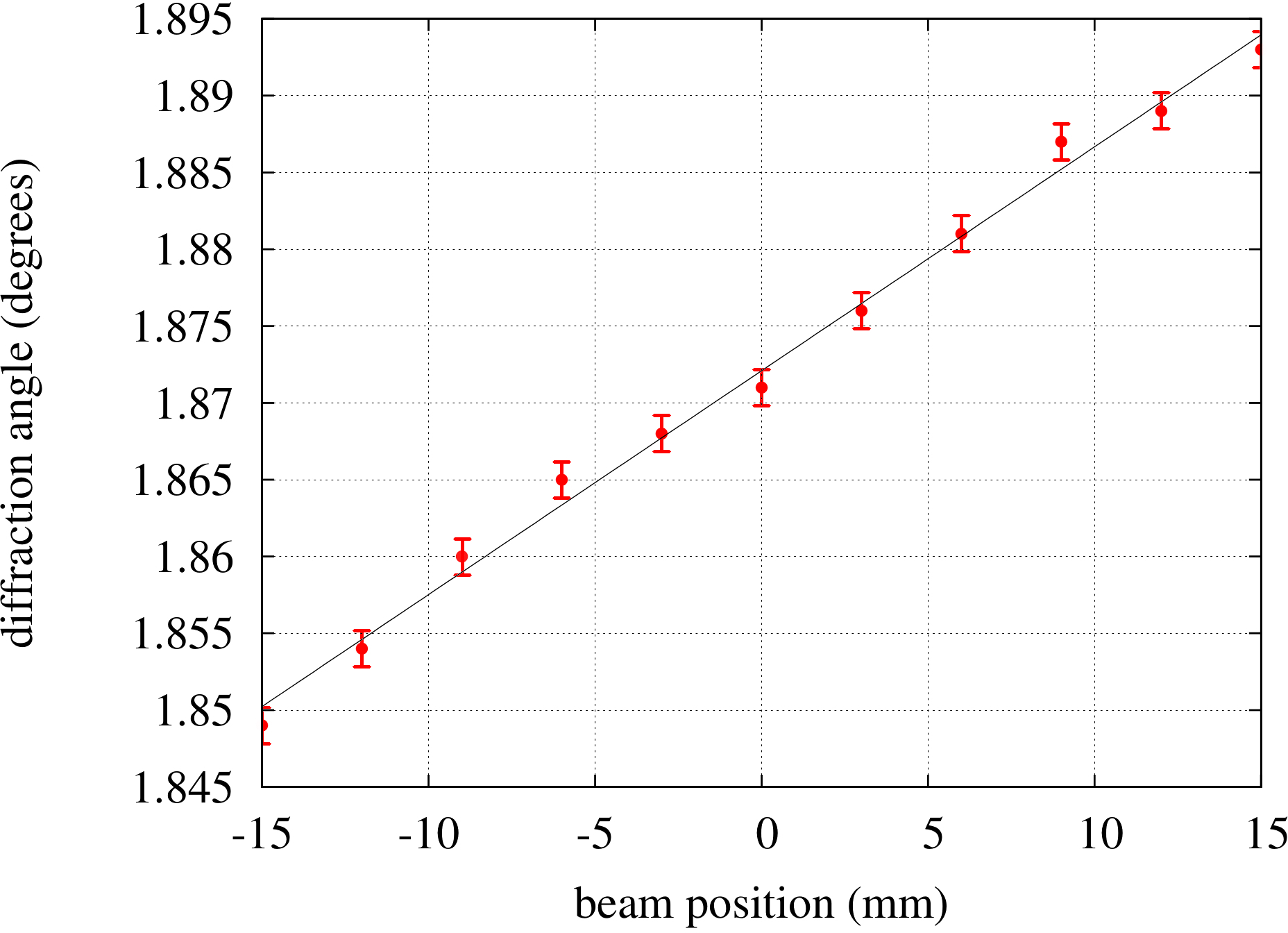}
     \caption{\footnotesize{Curvature radius estimation for the GaAs tile by exploiting the Tungsten k$\alpha$ line
     of the available X-ray tube.}}
     \label{curvradius}
\end{center}
\end{figure}

The crystals provided by {\sc cnr/imem} - Parma have been characterized in terms of efficiency and focusing 
effect before being mounted on the support of the Laue lens.
The presented tests were performed with one crystal sample of the received batch.
The curvature radius of the crystal was estimated through the X-ray fluorescence line of the X-ray tube anode
 available in the facility (Tungsten K$\alpha$ = 59.2~keV)  with the method already described in \cite{Liccardo2014}
 and it gives a value of $R_c$ = 39.4$\pm$1.5~m (see Fig.~\ref{curvradius}).

 \begin{figure}[!h]
 \begin{center}
   \includegraphics[scale=0.22]{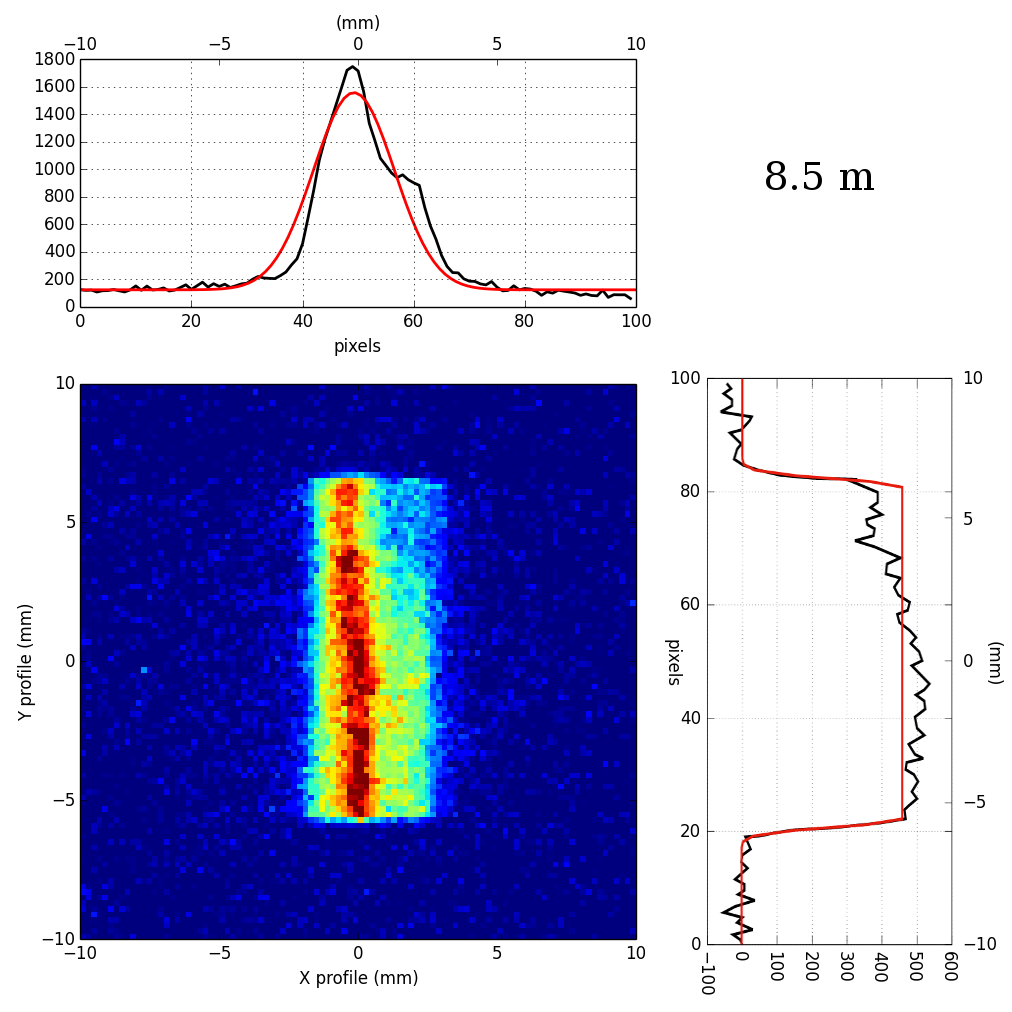}   
   \includegraphics[scale=0.22]{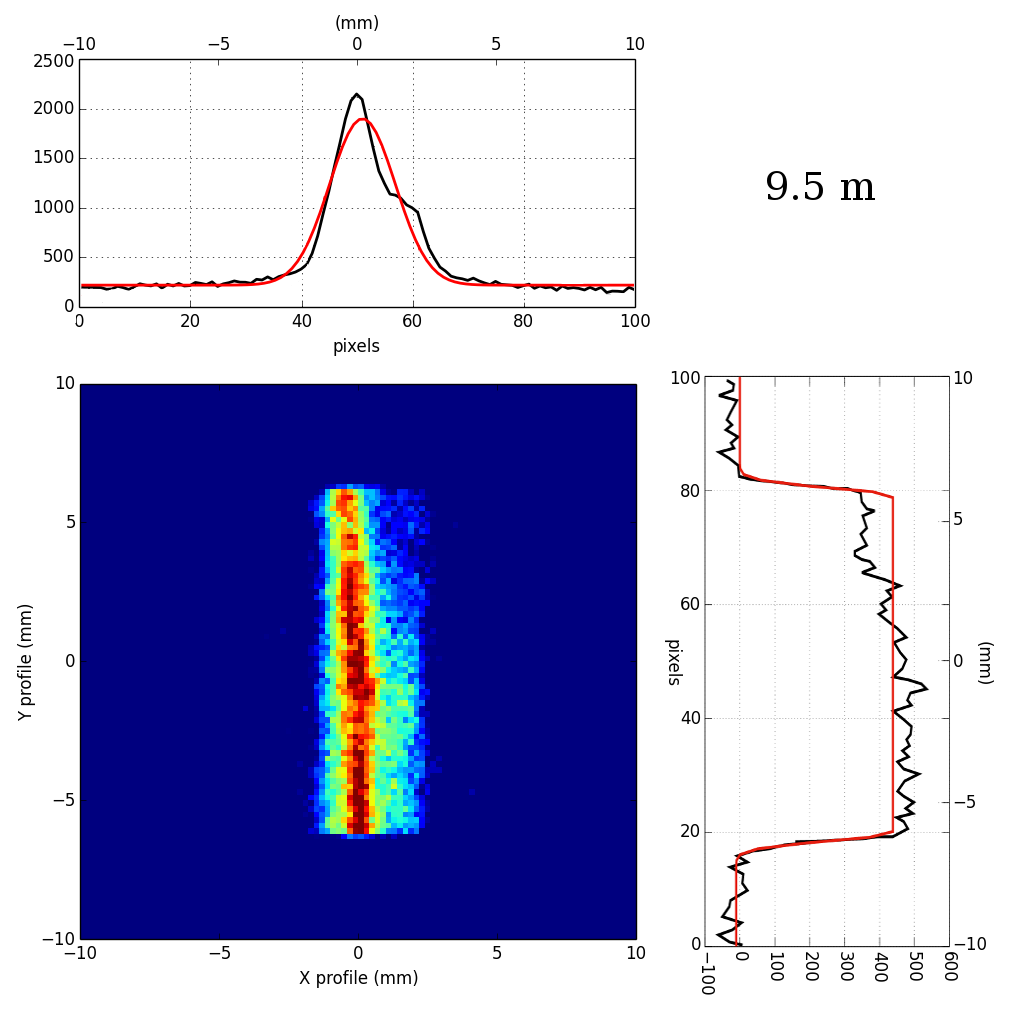}   
   \includegraphics[scale=0.22]{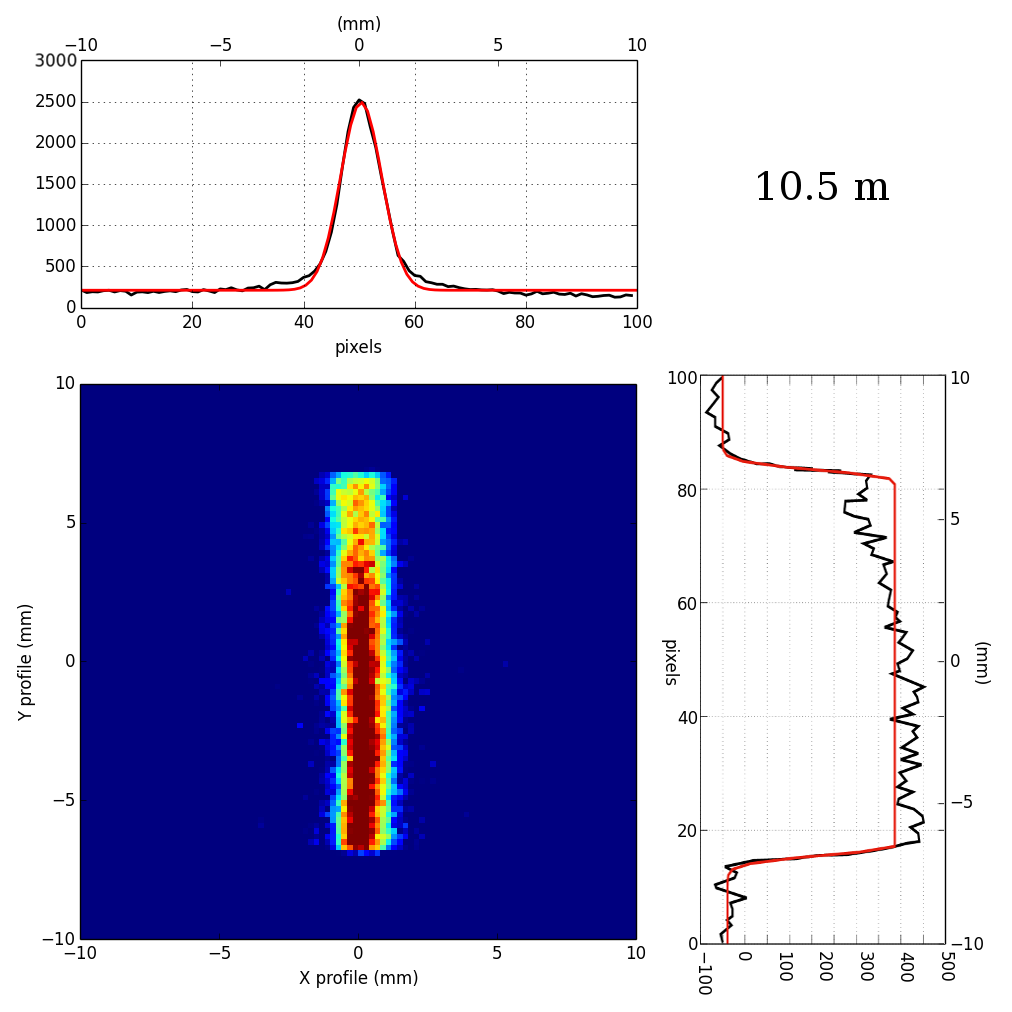}   
   \includegraphics[scale=0.22]{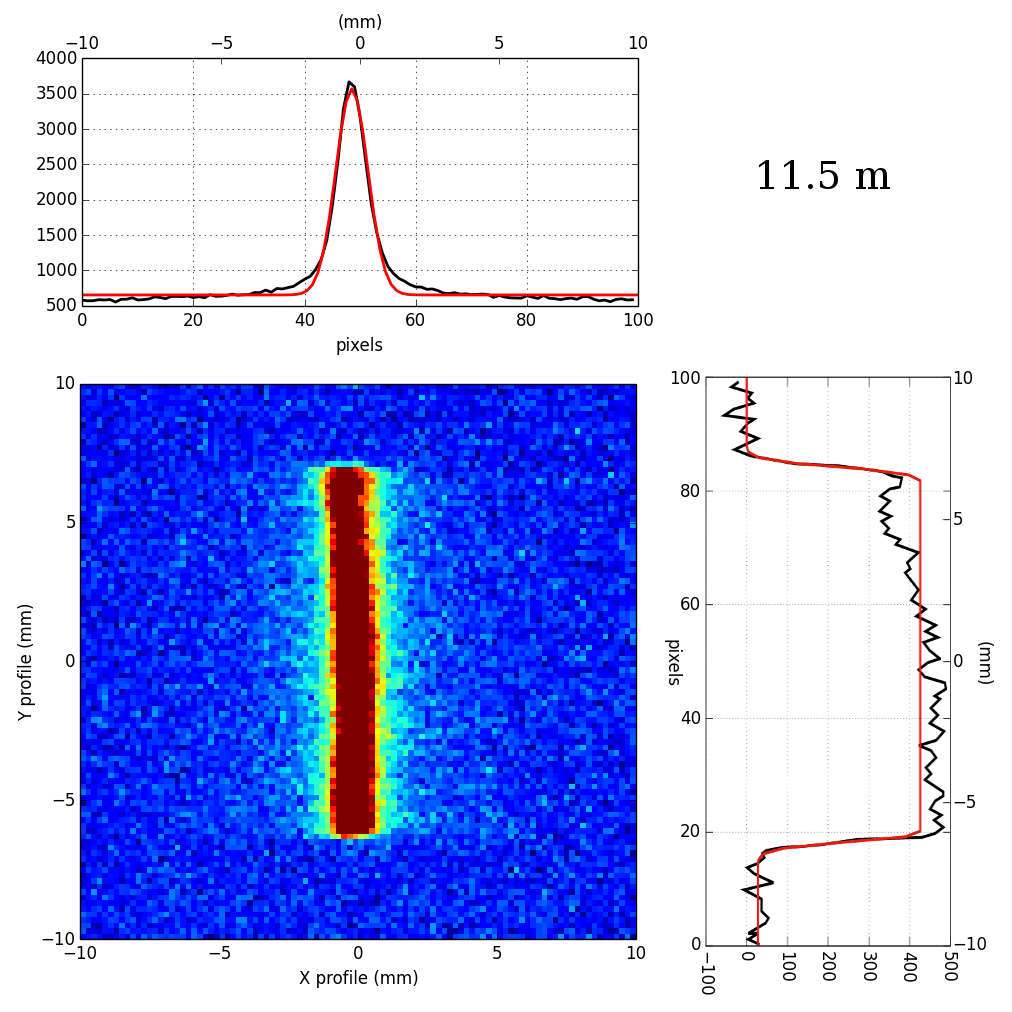}   
   \includegraphics[scale=0.22]{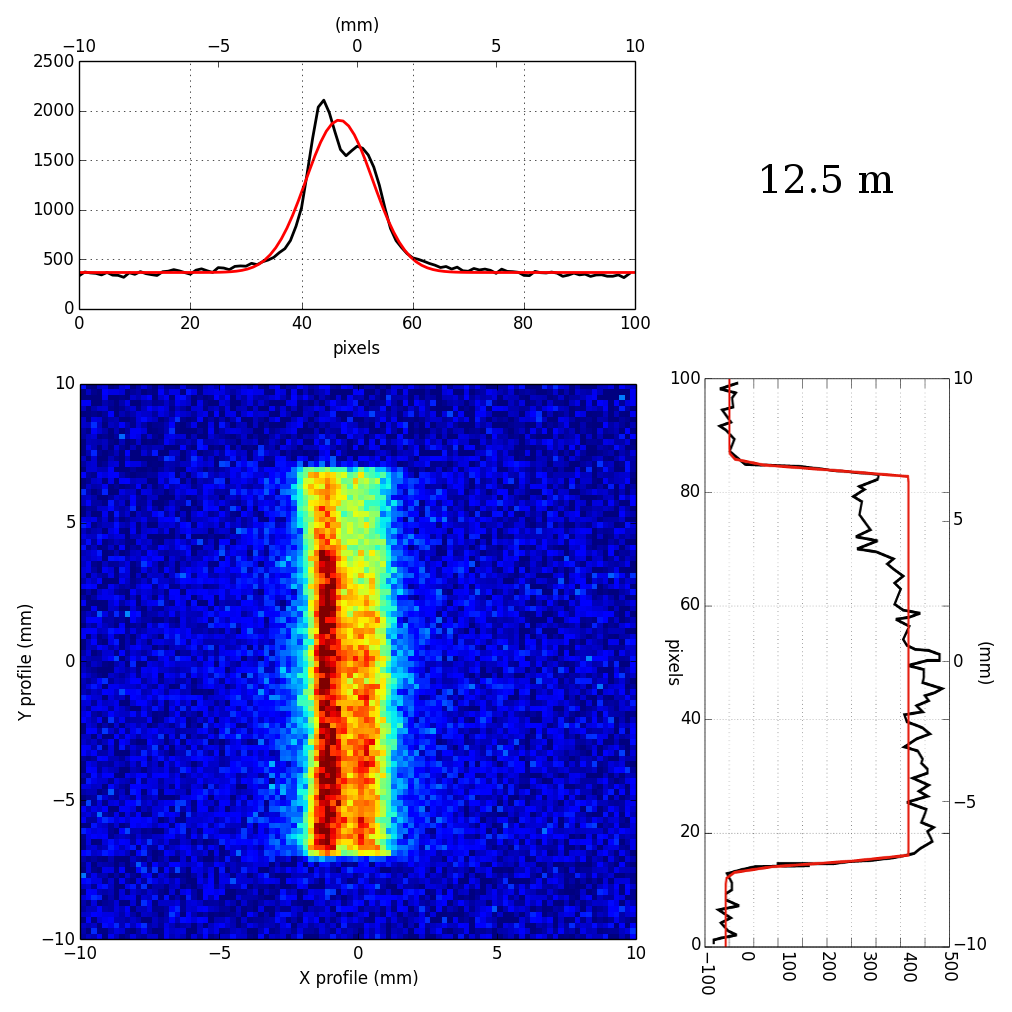}   
   \includegraphics[scale=0.22]{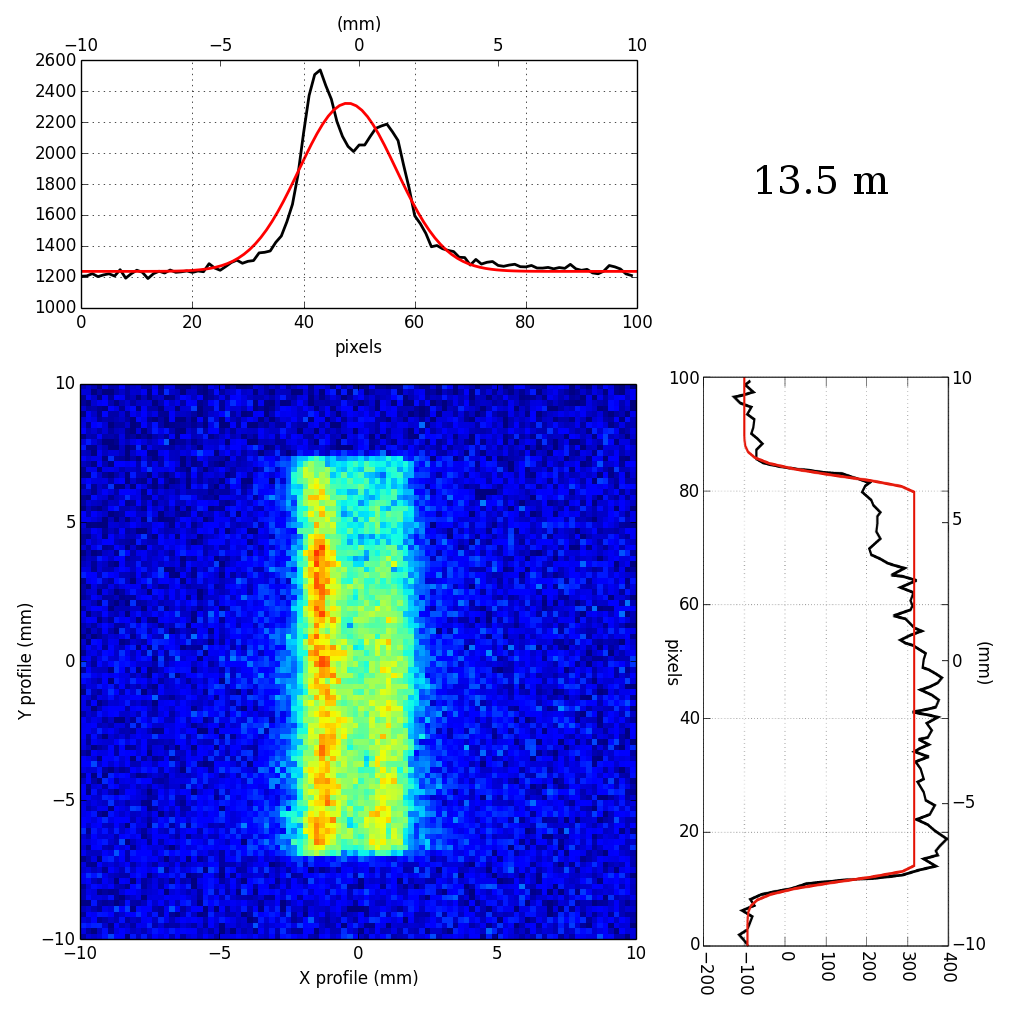}   
     \caption{\footnotesize{Diffracted beam measured at different distances crystal/detector, from the top left 
     to the bottom right 8.5, 9.5, 10.5, 11.5, 12.5, 13.5 m, respectively.}}
     \label{images:focaldistances}
 \end{center}
 \end{figure}

 
To test the described focusing effect, the beam size on the crystals was set to 20 $\times$ 10 mm$^2$, the longer size 
in the {\it x} coordinate, in order to better demonstrate the focusing effect along that direction.  
The image of the diffracted signal was studied with a total exposure time of 510 seconds. 
For comparison with the Monte Carlo simulations, the detector was progressively set at the same distances reported in 
Sec.~\ref{sec:mc} (8.5, 9.5, 10.5, 11.5, 12.5, 13.5 m). The results are shown in Fig.~\ref{images:focaldistances}, where 
it is evident that the different distance between crystal and detector results in a changing 
of both image shape and dimensions. 
The analysis of the {\it x} and {\it y} profile for each image of the diffracted beam 
acquired at different distances are also shown. 
As along the {\it x} profile the focusing effect of the GaAs mosaic crystal is present, 
it can be satisfactory described with a Gaussian function, plus a constant that represents 
the background counts which have not been subtracted.
The function adequately fit the data, mainly at distances close to 
the nominal distance $F_D$ = 11.40~m, while the fit becomes poor
far from the nominal position (e.g. at 8.5~m or 13.5~m).
This discordance between fit and data is caused by a not uniform  
curvature of the specific sample along its $x$ axis, which is 
emphasized particularly out of focus.
The {\it y} profile was instead modeled with the 
convolution  of a box function with a Gaussian and the width of the curve  
was measured at half of the peak value. 
In Tab.~\ref{table:fit}  the best fit values
of the measured {\sc fwhm} are reported and compared with those obtained 
from the Monte Carlo calculations (see Sec.\ref{mc}).

\begin{table}[!h]\renewcommand{\arraystretch}{1.}\addtolength{\tabcolsep}{0pt}
\begin{center}
\caption{\footnotesize Comparison between the {\sc fwhm} of the {\it x} and the {\it y} profiles for the acquired 
diffracted images, with the Monte Carlo simulations, at different distances between crystal and detector 
(R$_c$ = 40 m curvature radius).}
\label{table:fit}
\resizebox{10.5cm}{!}{
\begin{tabular}{ccc|cc}
     & \multicolumn{2}{c}{Experimental data} & \multicolumn{2}{c}{Monte Carlo}\\
\bottomrule
Crystal-detector    &  {\it x} {\sc fwhm}  &  {\it y}  {\sc fwhm}$^{(a)}$ &  {\it x} {\sc fwhm}   &   {\it y} {\sc fwhm} \\
  distance (m)      &             (mm)     &      (mm)            &      (mm)          & (mm)\\
  
  \hline
8.5                 &   4.95$\pm$0.21   & 12.33$\pm$0.32       & 4.45$\pm$0.07       &    13.16$\pm$0.10\\
9.5                 &   3.11$\pm$0.25   & 12.47$\pm$0.25       & 3.30$\pm$0.07       &    13.50$\pm$0.10\\
10.5                &   1.85$\pm$0.23   & 13.21$\pm$0.34       & 2.25$\pm$0.07       &    14.01$\pm$0.12\\
11.5                &   1.37$\pm$0.29   & 13.47$\pm$0.29       & 1.35$\pm$0.08       &    14.35$\pm$0.10\\
12.5                &   2.29$\pm$0.24   & 13.93$\pm$0.43       & 2.11$\pm$0.08       &    14.70$\pm$0.09\\
13.5                &   3.12$\pm$0.24   & 14.31$\pm$0.32       & 3.26$\pm$0.07       &    15.21$\pm$0.10\\
14.5                &   4.63$\pm$0.21   & 14.70$\pm$0.20       & 4.75$\pm$0.07       &    15.46$\pm$0.10\\
\midrule
\end{tabular}
}
\end{center}
$^{(a)}$  the {\sc fwhm} of the rectangular function is taken at half of the peak value of the curve. 
\end{table}

\begin{figure}[!h]
   \begin{center}
   \includegraphics[scale=0.35]{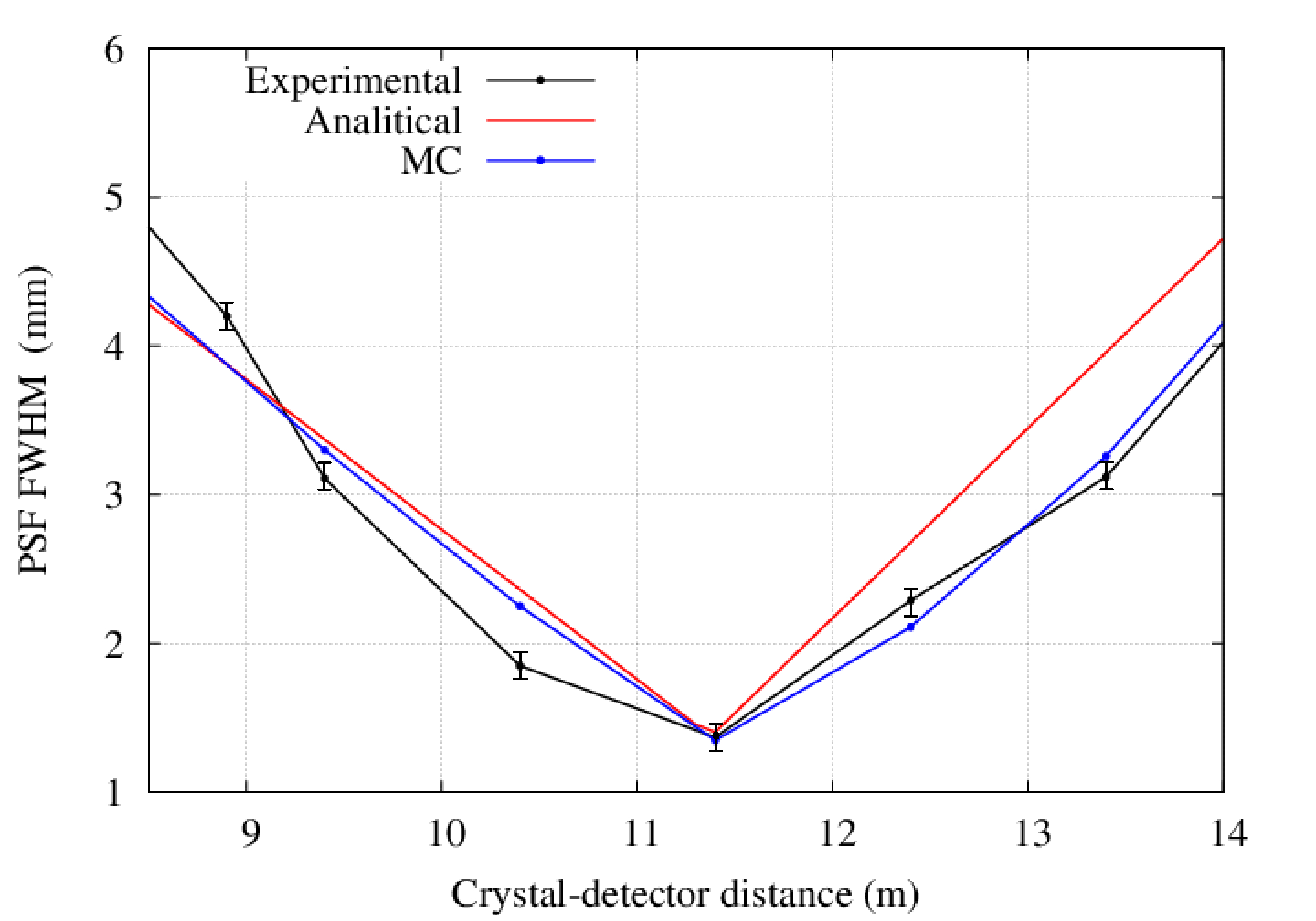}
     \caption{\footnotesize{Plot of the {\sc fwhm} of the {\it x} profile of the 
     diffracted image as function of the distance between crystal and detector, for 
     a GaAs (220) crystal with 39.9 $\pm$ 1.5 m curvature radius and fixed distance source-target of 26.40 m. 
     {\it Black points and line}: acquired data. {\it Blue points and line}: Monte Carlo simulations. 
     {\it Red line}: geometric-analytic estimation.}}
     \label{differentdistances}
 \end{center}
 \end{figure}

Along the  
{\it x} direction, where the 
focusing effect is expected,  we found 
a good agreement between experimental and Monte Carlo results.  Along the {\it y} coordinate,
no focusing effect is expected and the diffracted beam suffers a lengthening which is 
mainly due to the divergence of the beam. It is worth noting that the {\sc psf} in 
the {\it y} direction broadens with the increase of
the distance from the crystal. 
The agreement between experimental data 
and Monte Carlo is shown in Fig.~\ref{differentdistances}
where also the results based on the analytical calculations of Sec. \ref{div_effect} have been reported. 
Within the uncertainties, the agreement is satisfactory.

%

\section{Conclusions}
\label{sec:conclusions}
 
In this paper, a real focusing effect from a bent crystal of Gallium Arsenide (220) has been shown and discussed.
The sample, provided by {\sc cnr/imem} - Parma, was part of a batch of crystals 
bent by a lapping procedure. The requested curvature can be  
reproduced with a good precision (5\% uncertainty). For the {\sc laue} project a large number of tiles made of 
GaAs (220) and  Ge (111) are being mounted on a lens petal frame, in order to
build for the first time a Laue lens petal capable of operating over a broad energy band 
($\sim$ 90-300 keV). A systematic analysis of the diffracted profiles has been presented, using a 
Monte Carlo ray tracer with a set of tools capable of describing the focusing effect 
for both a parallel beam and a diverging beam.  Both the cases of flat crystals and bent crystals 
with 40 m curvature radius have been investigated.  All the processes of building and testing a Laue 
lens are made on ground while the lens is conceived for astrophysical applications.
For this reason a detailed study of the behavior in the case of 
a finite distance between the radiation source and the crystals cannot be avoided.
The good agreement found between experimental 
tests and Monte Carlo results give us the  confidence needed about the correctness of the Monte 
Carlo calculations also for the case of a parallel beam from an astrophysical source, 
that is the final goal of the {\sc laue} project.

\begin{acknowledgements}
The {\sc laue} team wish to tanks the Italian Space Agency
for the financial support under the contract I/068/09/0.
\end{acknowledgements}

\bibliography{references}

\begin{thebibliography}{10}

\bibitem{Lindquist68}
T.~R. {Lindquist} and W.~R. {Webber}, ``{A focusing X-ray telescope for use in
  the study of extraterrestrial X-ray sources in the energy range 20-140
  keV},'' {\em Canadian Journal of Physics Supplement}, vol.~46, p.~1103, 1968.

\bibitem{Smither95}
R.~K. {Smither}, P.~B. {Fernandez}, T.~{Graber}, P.~{von Ballmoos}, J.~{Naya},
  F.~{Albernhe}, G.~{Vedrenne}, and M.~{Faiz}, ``{Review of Crystal Diffraction
  and Its Application to Focusing Energetic Gamma Rays},'' {\em Experimental
  Astronomy}, vol.~6, pp.~47--56, Dec. 1995.

\bibitem{Halloin04}
H.~{Halloin}, P.~{von Ballmoos}, J.~{Evrard}, G.~K. {Skinner}, J.~M. {Alvarez},
  M.~{Hernanz}, N.~{Abrosimov}, P.~{Bastie}, B.~{Hamelin}, P.~{Jean},
  J.~{Kn{\"o}dlseder}, R.~K. {Smither}, and G.~{Vedrenne}, ``{Gamma-Ray
  Astronomy Starts to see CLAIRE: First Light for a Crystal Diffraction
  Telescope},'' in {\em 5th INTEGRAL Workshop on the INTEGRAL Universe}
  (V.~{Schoenfelder}, G.~{Lichti}, and C.~{Winkler}, eds.), vol.~552 of {\em
  ESA Special Publication}, p.~739, Oct. 2004.

\bibitem{Frontera07}
F.~{Frontera}, G.~{Loffredo}, A.~{Pisa}, L.~{Milani}, F.~{Nobili},
  N.~{Auricchio}, V.~{Carassiti}, F.~{Evangelisti}, L.~{Landi},
  S.~{Squerzanti}, K.~H. {Andersen}, P.~{Courtois}, L.~{Amati}, E.~{Caroli},
  G.~{Landini}, S.~{Silvestri}, J.~B. {Stephen}, J.~M. {Poulsen}, B.~{Negri},
  and G.~{Pareschi}, ``{Development status of a Laue lens project for gamma-ray
  astronomy},'' in {\em Proceedings of the SPIE}, vol.~6688, Sept. 2007.

\bibitem{Virgilli11a}
E.~{Virgilli}, F.~{Frontera}, V.~{Valsan}, V.~{Liccardo}, V.~{Carassiti},
  F.~{Evangelisti}, and S.~{Squerzanti}, ``{Laue lenses for hard x-/soft
  {$\gamma$}-rays: new prototype results},'' in {\em Proceedings of the SPIE},
  vol.~8147, id. 81471B 9 pp., 2011.

\bibitem{Zachariasen}
W.~H. Zachariasen, {\em Theory of X-ray Diffraction in Crystals}.
\newblock Wiley, 1945.

\bibitem{Lund2005}
N.~{Lund}, ``{An ``ESA-Affordable'' Laue-lens},'' {\em Experimental Astronomy},
  vol.~20, pp.~211--217, Dec. 2005.

\bibitem{Malgrange02}
Malgrange, ``X-ray propagation in distorted crystals: From dynamical to
  kinematical theory,'' {\em Crystal Research and Technology}, vol.~37, 2002.

\bibitem{ferrari13b}
C.~{Ferrari}, E.~{Buffagni}, E.~{Bonnini}, and D.~{Korytar}, ``{Curved focusing
  crystals for hard X-ray astronomy},'' {\em Crystallography Reports}, vol.~58,
  pp.~1058--1062, Dec. 2013.

\bibitem{Smither05}
R.~K. {Smither}, K.~A. {Saleem}, D.~E. {Roa}, M.~A. {Beno}, P.~{von Ballmoos},
  and G.~K. {Skinner}, ``{High diffraction efficiency, broadband, diffraction
  crystals for use in crystal diffraction lenses},'' {\em Experimental
  Astronomy}, vol.~20, pp.~201--210, Dec. 2005.

\bibitem{keitel99}
S.~Keitel, C.~Malgrange, T.~Niem{\"{o}}ller, and J.~R. Schneider,
  ``{Diffraction of 100 to 200 keV X-rays from an Si$_{1-x}$Ge$_{x}$ gradient
  crystal: comparison with results from dynamical theory},'' {\em Acta
  Crystallographica Section A}, vol.~55, pp.~855--863, 1999.

\bibitem{Kawata01}
H.~Kawata, M.~Sato, and Y.~Higashi, ``Improvements on water-cooled and doubly
  bent crystal monochromator for compton scattering experiments,'' {\em Nucl.
  Instrum. Meth. A}, vol.~467-468, pp.~404--408, 2001.

\bibitem{Bellucci11}
V.~{Bellucci}, R.~{Camattari}, V.~{Guidi}, I.~{Neri}, and N.~{Barri{\`e}re},
  ``{Self-standing bent silicon crystals for very high efficiency Laue lens},''
  {\em Experimental Astronomy}, vol.~31, pp.~45--58, Aug. 2011.

\bibitem{Camattari14}
R.~Camattari, E.~Dolcini, V.~Bellucci, A.~Mazzolari, and V.~Guidi, ``{High
  diffraction efficiency with hard X-rays through a thick silicon crystal bent
  by carbon fiber deposition},'' {\em Journal of Applied Crystallography},
  vol.~47, pp.~1762--1764, Oct 2014.

\bibitem{Buffagni11}
E.~{Buffagni}, C.~{Ferrari}, F.~{Rossi}, L.~{Marchini}, and A.~{Zappettini},
  ``{Preparation of bent crystals as high-efficiency optical elements for hard
  x-ray astronomy},'' {\em Optical Engineering}, vol.~51, p.~056501, May 2012.

\bibitem{virgilli13}
E.~{Virgilli}, F.~{Frontera}, V.~{Valsan}, V.~{Liccardo}, V.~{Carassiti},
  S.~{Squerzanti}, M.~{Statera}, M.~{Parise}, S.~{Chiozzi}, F.~{Evangelisti},
  E.~{Caroli}, J.~{Stephen}, N.~{Auricchio}, S.~{Silvestri}, A.~{Basili},
  F.~{Cassese}, L.~{Recanatesi}, V.~{Guidi}, V.~{Bellucci}, R.~{Camattari},
  C.~{Ferrari}, A.~{Zappettini}, E.~{Buffagni}, E.~{Bonnini}, M.~{Pecora},
  S.~{Mottini}, and B.~{Negri}, ``{The {\sc laue} project and its main
  results},'' in {\em Proceedings of the SPIE}, vol.~8861, id. 886106 17 pp.,
  2013.

\bibitem{ferrari14}
C.~{Ferrari}, E.~{Buffagni}, E.~{Bonnini}, and A.~{Zappettini}, ``{X-ray
  diffraction efficiency of bent GaAs mosaic crystals for the {\sc laue}
  project},'' {\em Optical Engineering}, vol.~53, p.~047104, Apr. 2014.

\bibitem{buffagni15}
E.~{Buffagni}, E.~{Bonnini}, C.~{Ferrari}, E.~{Virgilli}, and F.~{Frontera},
  ``{X-ray characterization of curved crystals for hard x-ray astronomy},'' in
  {\em Society of Photo-Optical Instrumentation Engineers (SPIE) Conference
  Series}, vol.~9510 of {\em Society of Photo-Optical Instrumentation Engineers
  (SPIE) Conference Series}, p.~6, May 2015.

\bibitem{camattari2012}
R.~{Camattari}, V.~{Guidi}, and I.~{Neri}, ``{Quasi-mosaicity as a tool for
  focusing hard x-rays},'' in {\em Society of Photo-Optical Instrumentation
  Engineers (SPIE) Conference Series}, vol.~8443 of {\em Society of
  Photo-Optical Instrumentation Engineers (SPIE) Conference Series}, p.~35,
  Sept. 2012.

\bibitem{Liccardo12}
V.~Liccardo, E.~Virgilli, F.~Frontera, and V.~Valsan, ``Characterization of
  bent crystals for laue lenses,'' in {\em Society of Photo-Optical
  Instrumentation Engineers (SPIE) Conference Series}, vol.~8443, 2012.

\bibitem{loffredo03}
G.~{Loffredo}, C.~{Pelliciari}, F.~{Frontera}, V.~{Carassiti}, S.~{Chiozzi},
  F.~{Evangelisti}, L.~{Landi}, M.~{Melchiorri}, S.~{Squerzanti}, S.~{Brandt},
  C.~{Budtz-Joergensen}, S.~{Laursen}, N.~{Lund}, J.~{Polny}, and N.~J.
  {Westergaard}, ``{X-ray facility for the ground calibration of the X-ray
  monitor JEM-X on board INTEGRAL},'' vol.~411, pp.~L239--L242, Nov. 2003.

\bibitem{larix}
E.Virgilli, {\em {\sc the larix facility}}, 2015.
\newblock \url{http://larixfacility.unife.it}.

\bibitem{Virgilli11b}
E.~Virgilli, V.~Liccardo, V.~Valsan, F.~Frontera, E.~Caroli, J.~Stephen,
  F.~Cassese, L.~Recanatesi, and M.~Pecora, ``The {\sc laue} project for broad
  band gamma-ray focusing lenses,'' in {\em Optics for EUV, X-Ray, and
  Gamma-Ray Astronomy V}, vol.~8147, 2011.

\bibitem{Liccardo2014}
V.~{Liccardo}, E.~{Virgilli}, F.~{Frontera}, V.~{Valsan}, E.~{Buffagni},
  C.~{Ferrari}, E.~{Bonnini}, A.~{Zappettini}, V.~{Guidi}, V.~{Bellucci}, and
  R.~{Camattari}, ``{Study and characterization of bent crystals for Laue
  lenses},'' {\em Experimental Astronomy}, vol.~38, pp.~401--416, Dec. 2014.

\end{thebibliography}
\bibliographystyle{ieeetr}

\end{document}